\documentclass[twocolumn]{aa}  

\usepackage{hyperref}
\usepackage{afterpage}
\usepackage{xcolor}
\usepackage{environ}
\NewEnviron{myresizeenv}{\resizebox{\linewidth}{!}{\BODY}}
\usepackage{arydshln}

\usepackage{graphicx}
\usepackage{txfonts}
\usepackage[figuresright]{rotating}
\usepackage{lscape}
\usepackage[sort&compress]{natbib}
\bibpunct{(}{)}{;}{a}{}{,}
\newcommand{\um}{$\mu$m}
\newcommand{\brgamma}{Br$\gamma$}

\def\arcsec{\hbox{$^{\prime\prime}$}}
\def\utw{\smash{\rlap{\lower5pt\hbox{$\sim$}}}}
\def\udtw{\smash{\rlap{\lower6pt\hbox{$\approx$}}}}

\def\Lsun{\hbox{\it L$_\odot$}}

\def\Msun{\hbox{\it M$_\odot$}}

\def\Teff{\hbox{\it T$_{\rm eff}$}}

\newcommand{\Ks}{{\it K$_{\rm s}$}}

\newcommand{\Aks}{{\it A$_{\rm K_{\rm s}}$}}

\newcommand{\Av}{{\it A$_{\rm V}$}}

\def\simgr{\mathrel{\hbox{\rlap{\hbox{\lower4pt\hbox{$\sim$}}}\hbox{$>$}}}}

\definecolor{myorange}{RGB}{230,100,0}
\definecolor{myred}{RGB}{180,0,0}
\definecolor{myblue}{RGB}{0,80,180}

\begin{document}
   \title{A catalog of \textit{Gaia} DR3 luminous blue variables  and candidate luminous blue variables}

   \subtitle{Limits of \textit{Gaia} DR3 light curves for identifying 
   luminous blue variables}
   \titlerunning{A catalog of \textit{Gaia} DR3 luminous blue variables  and candidate luminous blue variables}

\author{Maria Messineo\thanks{MM moved to INAF-OAPA in the spring 2026.}
          \inst{1,2,4}
          and
          Donald F. Figer
          \inst{3}
           }
\authorrunning{M. Messineo and D.F. Figer}

   \institute{Dipartimento di Fisica e Astronomia 
“Augusto Righi”, Alma Mater Studiorum, Università di Bologna, Via Gobetti 93/2, 
I-40129 Bologna, Italy\\
              \email{maria.messineo@inaf.it}
         \and
             INAF - Osservatorio di Astrofisica e Scienza dello Spazio 
di Bologna, Via Gobetti 93/3, I-40129 Bologna, Italy\\
  \and
Center for Detectors, Rochester Institute of Technology, 
54 Memorial Drive, Rochester, NY 14623, USA	\\
  \email{figer@cfd.rit.edu}
  \and
       INAF - Osservatorio Astronomico di Palermo, Piazza del Parlamento 1, 
       90134 Palermo, Italy\\
            }

   \date{Received September 15, 1996; accepted March 16, 1997}

 
\abstract
{ Galactic luminous blue variables (LBVs) are rare, 
evolved massive stars whose defining evidence is long-term 
spectroscopic and photometric instability, including 
S\,Doradus-type cycles, and, in exceptional cases, giant eruptions. 
 Candidate LBVs (cLBVs) share LBV-like spectra and/or circumstellar 
nebulae; however, they lack long-term observational datasets that would 
detect their long-term variability and confirm their classification as 
bona fide LBVs. Modern time-domain surveys provide homogeneous light 
curves for many Galactic massive stars, but each survey has its own 
limitations in terms of depth, temporal coverage, and observing cadence.
}
{ Using confirmed Galactic LBVs as a control sample, 
we aim to test what \textit{Gaia} DR3 and complementary photometric surveys 
can and cannot establish about LBV-type variability.
}
{ We assembled a vetted sample of 75 Galactic LBVs and 
cLBVs with near- and mid-infrared photometry from 2MASS, MSX, 
WISE, GLIMPSE, and MIPSGAL. The infrared material was used to 
validate identifications, evaluate source confusion, 
and place objects in a circumstellar-dust context. 
We then constructed light curves from \textit{Gaia} DR3, the 
Bochum survey, the All-Sky Automated Survey for SuperNovae (ASAS\,SN), and 
the Zwicky Transient Facility (ZTF), 
and analyzed large-amplitude and periodic cases with 
Lomb-Scargle (LS) and phase dispersion minimization (PDM) methods.
}
{ \textit{Gaia} DR3 matches are available for 67\% of the sample and
76\% are reported  as variable. 
We identify high-amplitude variables and 
periodic systems that may be binaries rather than classical LBVs.
Six confirmed LBVs and five candidates
show variations  larger than $\approx$ 0.6 mag in the \textit{Gaia} $RP$ filter. 
Their light curves are examined and three  
periodic variables are identified. 
The light curves of  OMN2000\,LS1 and DWC2011\,140 
show primary and secondary minima, 
strongly suggesting eclipsing binary systems.
OMN2000\,LS1 has a well-constrained period of 80.29 d, 
and DWC2011\,140 of 95.07 d.}
{}
   \keywords{Stars: massive --
                Stars: emission-line, Be --
                Stars: variables: general --
                binaries: general --
                binaries: eclipsing --
                catalogs
               }

   \maketitle
%

\section{Introduction}

The evolutionary path of massive stars is not well 
deciphered. In addition to initial masses and metallicity 
(hence opacity of stellar atmospheres),  rotation, mass-loss, magnetic field, 
and binarity are key parameters in understanding 
their evolution and final fate. 
Stars from $\approx $8--40  \Msun\  
(or $\approx $8--25  \Msun\ when rotation is added) 
become  red supergiants  \citep[RSGs, e.g.,][]{limongi17}, then they may explode or 
move back to hotter temperatures as transitional objects (post-RSGs);
stars more massive than that do not become RSGs,
but directly evolve blueward - from blue supergiants (BSG),
to transitional objects and Wolf--Rayet (WR) stars.
Transitional evolved massive stars include LBVs, B[e] supergiants (B[e]SGs), 
yellow supergiants and yellow hypergiants, BSGs, 
and Of/WN stars. Their relative numbers and 
environments constrain the effects of mass loss, rotation, metallicity, 
and binarity on massive-star evolution.
As defined by 
\citet{kraus19}, B[e]SGs show evidence of a dusty disk—often 
interpreted as supporting binarity—such as CO emission or a 
hot dust component.  
For B[e]SGs,  $\log(L/\Lsun)$ ranges 
from 4.0--6.1 \citep{smith15,kraus19},
while LBVs are generally brighter: 
their  log(L/\Lsun) ranges from $4.5$ to 6.7
\citep[e.g.,][]{smith19,humphreys16,smith15,vangenderen01}.

Modern time-domain surveys, including \textit{Gaia} DR3, 
ASAS\,SN, ZTF, and Transiting Exoplanet Survey Satellite (TESS), 
now provide homogeneous light curves for 
large samples of Galactic massive stars. However, optical 
surveys remain insensitive to many highly reddened LBVs 
embedded in the Galactic plane, and their relatively 
short temporal baselines cannot capture the decades-long 
evolution that defines classical LBVs.

The LBVs are not defined by a single 
photometric amplitude threshold. They are high-luminosity 
evolved massive stars with dense unstable winds, 
emission-line spectra, and documented long-term spectroscopic 
or photometric changes, most characteristically S\,Doradus cycles 
or giant eruptions. In contrast, candidate LBVs (cLBVs) have spectra, 
luminosities, nebulae, or infrared excesses that resemble those 
of LBVs, but lack decisive historical evidence for 
LBV-type variability. This distinction is central to the 
present paper: short-baseline surveys can reveal variability, 
periodicity, and possible binarity, but
generally cannot replace the long historical record needed 
to confirm an LBV.

The environment 
provides a critical context for understanding 
their history of mass-loss and their evolutionary state.
\citet{smith15} concluded that LBVs often reside 
in relative isolation.
Other studies argue that observational 
biases and incomplete 
cluster memberships may affect the statistics
\citep[e.g., ][]{humphreys16,martin25}. 
Some studies argue that the initial mass (or luminosity) 
is the key parameter for the spatial distribution of stars,
with  massive LBVs being more clustered 
\citep{humphreys16,kostenkov25}, while others suggest 
that dusty LBVs could be post-supernova kicked stars
\citep[e.g., ][]{deman24}.
Questions that arise are whether there are two populations of LBVs 
and if the faintest are  binaries.  

For a long time, LBVs were considered 
a phenomenon of single stars
\citep[e.g.,][]{conti95,nota95,figer97,clark05}. 
However, recent scenarios also suggest that many LBVs are 
products of binary or multiple systems \citep[e.g.,][]{smith15}. 
Recent studies  with interferometry  
indicate a high binary fraction of $\sim 70\%$ 
among LBVs \citep{mahy22}, supporting the role of binary interactions 
in explaining their isolation. Radial-velocity periodicity is, however, 
rarely detected,  and binaries do not 
seem to be the cause of typical S\,Doradus cycles, or outbursts 
\citep{lobel25,mahy22}.\\

The practical question addressed here is, therefore, 
limited but important: 
whether LBV-like variability is detectable in the 2.8 yr \textit{Gaia} DR3 baseline, 
and whether cLBVs selected from the literature show variability 
inconsistent with simple LBV classification. 
We describe 
the photometric variations featuring LBV stars
in Sec.  \ref{oscilla}, construct a vetted catalog of Galactic LBVs and cLBVs
(Sect. \ref{sample}), 
and retain infrared photometry as supporting information for 
source validation (Sect. \ref{infrared}) and
comparison with B[e]SG-like objects in the
circumstellar-dust context.
We then use Gaia DR3, Bochum, ASAS\,SN, 
and ZTF light curves 
to measure amplitudes, identify high-amplitude variables 
(Sect. \ref{variability.section}), 
and search for periodicity.
The three resulting periodic cLBVs are discussed as likely binary 
candidates rather than as newly confirmed classical LBVs 
(Sect. \ref{sect.periodic}).
A summary is given in Sect. \ref{summary.sect}.

\section{Photometric variations of known LBVs}
\label{oscilla}
Bona fide Galactic LBVs provide the empirical reference 
sample for interpreting the modern survey light curves. 
Because LBV variability occurs on multiple timescales, 
no single amplitude threshold can be used as a classification rule. 
Giant eruptions are rare, century-scale events with 
visual brightenings of several magnitudes. 
S\,Doradus cycles are more common but still require years to decades 
of monitoring and combine visual-brightness changes with temperature 
and spectral changes at approximately constant bolometric luminosity. 
Shorter oscillations and microvariability occur on days-to-years 
timescales and can be detected in modern surveys, but these alone 
are not sufficient to confirm an LBV.
The historical thresholds used by van Genderen and 
Sterken\footnote{LBVs confirmed by Dr. van Genderen 
show variations larger than $\approx 1$ mag
\citep[][]{vangenderen97,vangenderen92,vangenderen97b},  
while those confirmed by Dr. Sterken  have variations 
as small as $\approx 0.4$ mag \citep{sterken96,sterken08}.} 
should, therefore, be regarded as empirical guidance rather than formal 
definitions. Long-term, large-amplitude changes helped establish 
many bona fide LBVs, but smaller-amplitude objects can also be 
LBVs when spectroscopy, color behavior, and historical variability 
support that interpretation.  

The most striking type of variations are giant eruptions, 
documented for P\,Cygni and $\eta$\,Carinae  
\citep[e.g.,][]{degroot88,vangenderen01},
during which a star can brighten by several magnitudes 
and eject large amounts of mass. 
These rare events occur on timescales of centuries or longer.\\
The defining property of the S\,Doradus  
cycle  \citep[e.g., HR\,Carinae, AG\,Carinae, ][]{vangenderen97,vangenderen97b} 
is the anticorrelation between brightness 
and color (or effective temperature) changes: 
LBVs become  redder and brighter (maximum light) 
and bluer and fainter  (minimum light), 
corresponding to visual 
amplitudes of typically  0.2–2.5 mag and periods of 10-40 yr, 
while maintaining nearly constant bolometric 
luminosity. 
Spectral types shift accordingly, 
from hot B-type supergiants  or Of/WN states in the minimum state to cooler 
A–F types in the maximum state.\\
LBVs also display oscillations (in colors and magnitudes) 
of about half a magnitude on a timescale from months to a few  
years along with the S\,Doradus-type cycles \citep{humphreys94}.
Both LBVs and $\alpha$ Cygni variables\footnote{ $\alpha$ 
Cygni variables are non-radially pulsating luminous 
supergiants with irregular or multiperiodic low-amplitude brightness 
variations; unlike LBV S\,Doradus cycles, they do not show 
the same systematic long-term color-temperature excursions.}
may display  microvariability
on shorter timescales of days to months, 
but in LBVs their amplitudes are larger (typically 0.1--0.25 mag)  
and occur  near the minimum and maximum light curves 
 \citep[][]{vangenderen97b,vangenderen92,vangenderen01,sterken08}.
Recent photometric monitoring with 
TESS of about 40 LBVs across the populations of the Galactic and Magellanic Clouds (MCs) 
confirm  short-term microvariability  on timescales  of days to months,
which is uncorrelated with stellar parameters \citep{spejcher25}. 
Extraordinarily, large semi-amplitudes are reported for
 AG\,Carinae   and  WRAY\,19-46 \citep{spejcher25}.
A  suitable light curve of P\,Cygni, covering from 1949 to 1955 is 
published  by \citet{humphreys94}; it shows 
oscillations with large amplitudes (0.5--1.0 mag).\\ 
Quasi-periodic variability is observed in some LBVs.
Between 1994--1998,  WRAY\,751 
seems in a low brightness state (before the brightening of 2001--2008), 
and the light curve shows quasi-periodic (of about 1 yr) variations 
\citep[full amplitude $\approx 0.5$ mag,][]{sterken08}; basically, this part of the 
light curve looks similar to that of a semiregular late-type star.\\
In the historical light curve of the great eruption
of  $\eta$\,Carinae  in 1837, three large  oscillations (with amplitudes $>0.5$ mag) 
are observed after the slow luminosity decline post-burst \citep{smith11,smith11b}.\\ 
After the burst of 1985 and consequent decline, between 1985 and 1990, 
AG\,Carinae  had quasiperiodic (300 d) oscillations with variable 
amplitude ($\approx 0.4$ mag) \citep{humphreys94}.

In conclusion, many modes of pulsations are at play in LBVs, 
with  different timescales and amplitudes  \citep[e.g.,][]{levesque25,guzik12}.
While P\,Cygni has microvariations with a period (P) of 17.3 d 
\citep[amplitudes of 0.1–-0.3 mag,][]{humphreys94, vangenderen92}
based on BRITE satellite data, its oscillations (pressure waves) have 
multiperiodicity (P=100, 175, 500 d), long-term photometry 
(gravity waves or subsurface convection) have long periods (P=900--1000 d), 
and the S Doradus phase has P=7.3 yr (amplitude 0.1 mag) \citep{elliott22}.

\section{Sample}
\label{sample}
In the literature, several recent reviews 
\citep[e.g.,][]{prasoon25,smith25}
and compilations of catalogs of LBVs and 
cLBVs have been reported. For example, in their Table 1,
\citet{richardson18} present  the names and coordinates 
of 36 Galactic LBVs\footnote{We actually 
found this catalog after compiling our list. 
All Galactic objects, except for 
two, are included in our compilation. HD\,148937 
is identified as a magnetic 
O-type star by \citet{lim24}; 
and MWC\,1015, a member of Berkeley\,87, has been reported to change from a late-type 
to a Be-type star \citep{polcaro06}.}. Similarly, \citet{weis20} list 15 confirmed 
LBVs and 32 candidates in their Table 2. 

We recompiled a list of LBVs and cLBVs from the literature.
We considered the list of 35 stars, 12 bona fide LBVs and 23 cLBVs 
given by \citet{clark05}, the list of six new candidates detected by
\citet{wachter10}\footnote{\citet{wachter10} 
list ten LBVs and cLBVs, but only six
are newly detected in their survey.}, the list of 16 bona fide Galactic LBVs 
by \citet{kniazev15} to which two  bona fide stars were added by 
\citet{gvaramadze17}, 
the list of 36 new cLBVs discovered with Spitzer 
and WISE data by  \citet{gvaramadze17}\footnote{The 36  cLBVs are a subcatalog of 
those stars discovered with Spitzer and WISE data for which the authors
had spectroscopic information 
\citep[and references therein,][]{gvaramadze17}. }, 
the two  candidates (2MASS\,J161517$-$505219 and 2MASS\,J115443$-$631331)
found with SALT spectra by \citet{kniazev15b}, with 
2MASS\,J161517$-$505219 being spectroscopically analyzed by \citet{figer20}.
The list of 25 stars with Gaia DR2 distances of
\citet{smith19} adds one new candidate (SBW1).
\citet{messineo14} report  another new cLBV 
in the G23.3$-$0.3 giant molecular complex, 
and \citet{messineo15}   a new cLBV in the cluster Cl\,1813-178.
More recently, eight candidates are reported toward the central molecular 
zone  by \citet{clark21}.
The collected stars are labeled as LBV or cLBVs as in recent literature.
WRAY\,17-96 is reported as a cLBV by \citet{clark05}; however,
because of the large photometric  variation reported by  \citet{buemi11},
we changed the status from cLBV to bona fide LBV.
 From a collection of about 90 stars, we removed
those very recently listed as BSGs, blue hypergiants (BHG),
Wolf-Rayet stars (WNL), post-asymptotic giant branch star
 (post-AGB), and Ofpe/WN9 stars by \citet{bernini26}.
In conclusion, the collected list of  75 stars includes  17 bona fide LBVs 
and  58 cLBVs. 

The collected stars are reported to show  emission lines
typical of LBVs in their spectra,  to be mostly 
surrounded by nebulae and brighter than $10^5$ \Lsun.
The temperatures and luminosities of LBVs,
available  from the literature, are listed in the Appendix
and are shown  in the diagram of Fig.\ \ref{fig.lum-teff}, with 
$\eta$\,Carinae  reaching $10^{6.7}$ \Lsun.
As shown by  \citet{smith19}, there is a large overlap 
on the faint end with B[e] stars, and
WS1 is a faint  highly-probable LBV star which has a   
low luminosity of $10^{4.5}$ \Lsun\ \citep{smith19}.

\begin{figure}[!]
\begin{center}
\resizebox{.99\hsize}{!}{\includegraphics[angle=0]{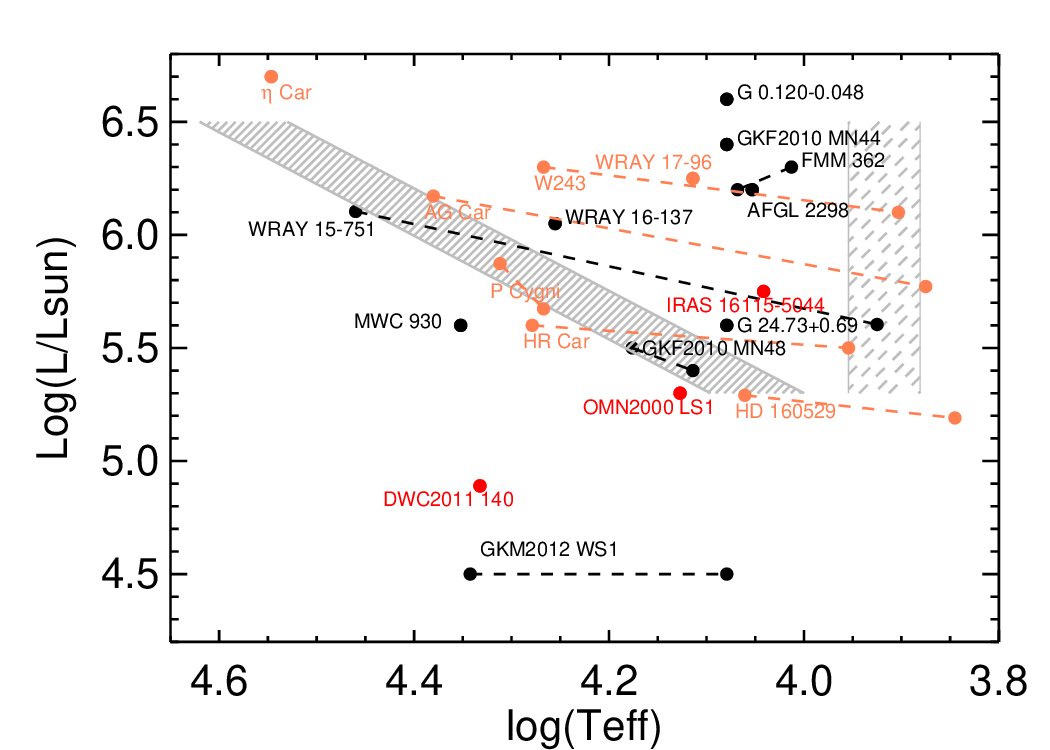}}
\end{center}
\caption{\label{fig.lum-teff} 
Luminosity vs. \Teff\ values of bona fide LBVs  (filled black circles);
the luminosities and \Teff\  values are taken
from the literature and the references are 
listed in the Appendix \ref{appendixlum}.
Those bona fide LBVs known to be binaries are colored in orange 
\citep[][and references therein]{mahy22,lobel25}, 
as listed in Table \ref{lbvlist}.
The cLBVs  OMN2000\,LS1,  IRAS\,16115$-$5044, and DWC2011\,140,
which are largely discussed in this work and for which 
luminosities are available in the Appendix \ref{appendixlum},
are over-plotted  with red-filled circles. 
OMN2000\,LS1 and DWC2011\,140 are newly reported 
as eclipsing binaries in this work.
The hot edge of the LBV minimum strip (continuous line) and 
the outburst region  (region enclosed between the two long-dashed
vertical lines) are taken from \citet{smith04}.  }
\end{figure}

\section{Observational data}
\label{infrared}

\subsection{Available infrared photometry}
Near- and mid-infrared photometry was collected for two main reasons.
First, it provides robust source identifications for highly reddened 
Galactic-plane objects, where optical counterparts can be faint or confused. 
Second, mid-infrared colors help trace 
circumstellar dust and nebular emission, 
which are relevant for distinguishing dusty cLBVs from 
disky B[e]SG-like objects and other evolved 
emission-line stars.
We used the near-infrared Two Micron All Sky Survey 
(2MASS)\footnote{Two Micron All Sky Survey \citep{skrutskie06}.} 
to correctly reidentify and verify the central stars  
\citep[e.g.,][]{gvaramadze12}. \\
Mid-infrared data were collected  
from the Midcourse Space Experiment (MSX), 
Wide-field Infrared Survey Explorer (WISE) surveys, 
and from the Spitzer Space Telescope surveys 
of the inner Galactic plane\footnote{ The 
Midcourse Space Experiment (MSX) is a survey of the Galactic plane 
from 8.28 \um\ to 21.3 \um\ \citep{egan03};
the Wide-field Infrared Survey Explorer (WISE) is an all-sky survey 
from 3.4 to 22 \um\ \citep{wright10};
the Galactic Legacy Infrared Mid-Plane Survey 
Extraordinaire, GLIMPSE, acquired images from 3.6 to 8.0 \um\ 
with the Spitzer/IRAC photometer \citep[][]{churchwell09}; 
the Multiband Imaging Photometer for Spitzer Galactic Plane Survey, 
MIPSGAL, observed at 24\,$\mu$m and 70\,$\mu$m \citep{gutermuth15}.
}.
These surveys are designed to trace infrared emission from warm 
dust in circumstellar nebulae, star-forming regions, and evolved stars.\\ 
The detailed catalog references and the full photometric compilation 
are provided in Appendix\,C.

\subsection{Gaia DR3 data}
Fifty stars (14 LBVs and 36 cLBVs)  have a  corresponding  counterpart
in the  Gaia DR3 catalog, as  listed in the Appendix \ref{amplitudes.table}.
For automatic Gaia positional matches, we preferred Gaia matches verified 
in the SIMBAD database and Gaia matches corresponding to the
2MASS counterparts; each match was inspected on the Gaia Archive
Single Object map\footnote{$https://gea.esac.esa.int/archive/$}. 

\subsubsection{XP spectra}

Forty-two stars have low-resolution spectra 
from the Blue Photometer (BP) 
and the Red Photometer (RP) of Gaia. 
The combined BP and RP data are referred to as XP spectra (or BPRP spectra). 
These low-resolution spectra ($R \approx 50$) cover approximately 330--1050\,nm 
\citep{deangeli23} and are useful primarily 
for broad spectral signatures, such as strong 
\ion{H\kern-0.2em}{$\alpha$} emission or broad blended features.
 A few examples are shown in Fig. 2. 
At R $\approx 50$, the narrow P\,Cygni profiles and 
the weak emission lines are not resolved, so the XP spectra 
are treated as supporting evidence rather than as a substitute 
for ground-based spectroscopy.
In more than half of the sample (25 out of 42), XP spectra  
show the \ion{H\kern-0.2em}{$\alpha$}  line in emission;
most of the spectra appear otherwise featureless
or with unresolved broad bumps.
Strong \ion{H\kern-0.2em}{$\alpha$}  lines are seen only in the 
spectra of nine stars 
(HR\,Carinae, WS1, P\,Cygni, Hen\,3-519, WRAY\,15-906, 
MN7, HD\,316285, MWC\,314, and  MWC\,349A). \\
HD\,326823 and MWC\,314 have a notable broad emission feature,
which is identified with Ca\,{\sc ii} in emission
compared to high-resolution spectra
\citep{borges01,miroshnichenko98}.\\
The feature around 710 nm in the spectrum
of Hen\,3-519, WS1, HD\,326823, and MN112  is due to \ion{He}{i} at 706.5  nm,
as seen in the high-resolution spectra of \citet{smith94b},
\citet{kniazev15}, \citet{borges01}, and \citet{kostenkov20}, respectively.
This weak feature also appears in the spectra of  HR\,Carinae, AG\,Carinae,
MN1, WRAY\,15-906, and HD\,316285.\\
 The small double bump that occurs between 920--960 nm
in the spectra of WS1,  WRAY\,15-906, MN7, WS2, 2MASS\,J17470940$-$2849235,
 G26.47+0.02, and MWC\,349A is most likely due to emission of 
$H$ lines of the Paschen series. Indeed,
$\eta$\,Carinae and HD\,316285 show strong $P_\zeta$ (922.9 nm) and
$P_\epsilon$ (954.6 nm) lines  \citep{hillier01,hillier98}.

\begin{figure*}[!]
\begin{centering}
\resizebox{0.33\hsize}{!}{\includegraphics[angle=0]{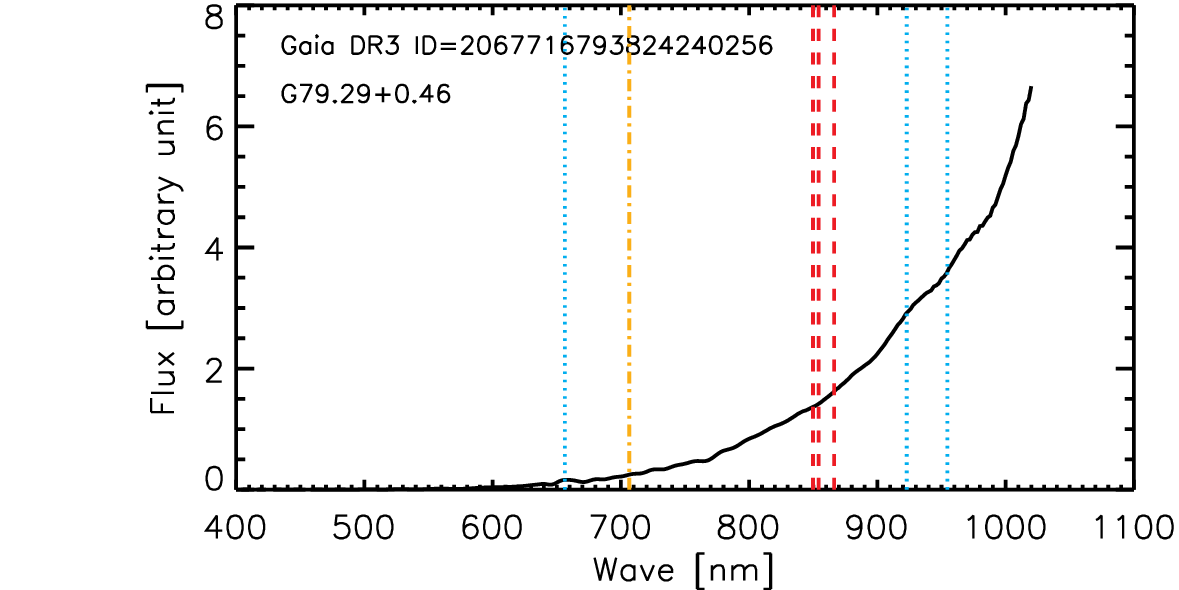}}
\resizebox{0.33\hsize}{!}{\includegraphics[angle=0]{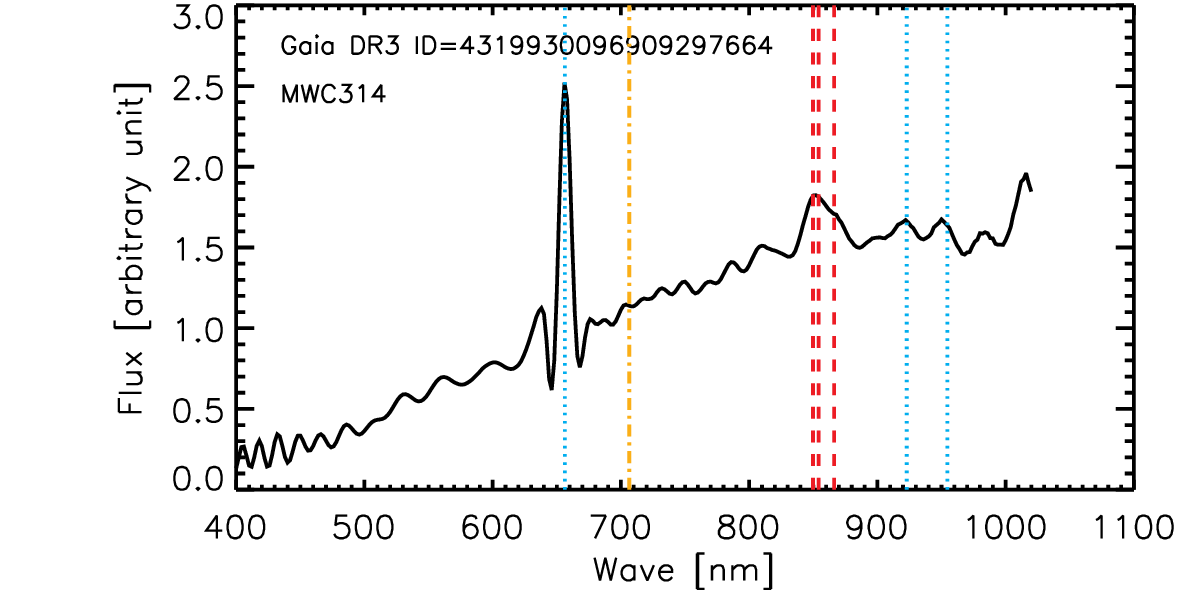}}
\resizebox{0.33\hsize}{!}{\includegraphics[angle=0]{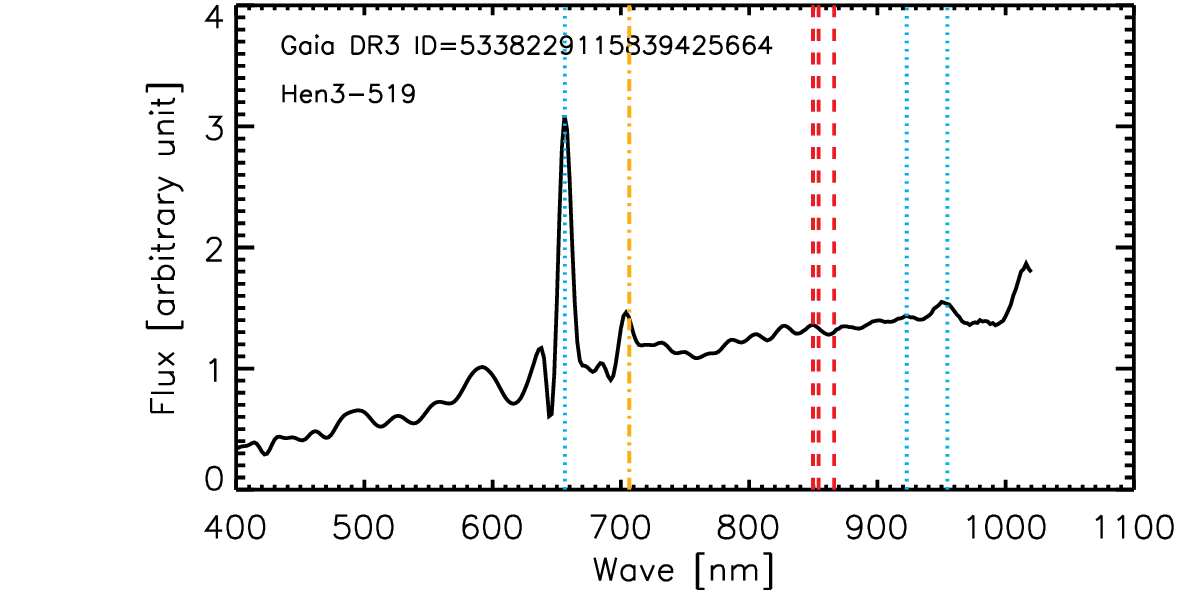}}
\end{centering}
\caption{\label{bprpfig_lbv} 
 Examples of Gaia DR3 XP spectra of known LBVs and cLBVs.
{\it Left:} Spectrum of G79.29+0.46.
{\it Middle:} Spectrum of MWC\,314.
{\it Right:} Spectrum of Hen\,3--519.
The position of the \ion{He}{i} line at 706.5~nm, which appears in a few spectra, 
is indicated by a yellow dot--dashed line.
The three strongest hydrogen lines identifiable in some spectra,
\ion{H\kern-0.2em}{$\alpha$}  (656.3~nm), $P_\epsilon$ (954.6~nm),
and $P_\zeta$ (922.9~nm), are marked by vertical cyan dotted lines.
The \ion{Ca}{ii} triplet (849.8, 854.2, and 866.2~nm) 
is marked by red dashed lines.
}
\end{figure*}

\subsubsection{Gaia variables}

There are  38 stars (out of 50) classified as variables in Gaia DR3.
Gaia has sparse epochs 
\citep[30–100 epochs over 2.8 yr,][]{eyer23}. 
The light curves show some outliers, and
those were discarded from the light curves imposing
that each data point should not deviate more than 
three times the standard deviation of
neighboring data points in time  (within 365 days).
The amplitudes of the variations were estimated as the 
difference between the
minimum and maximum magnitudes. This method is more conservative
than using as amplitudes the  default Gaia 
trimmed-magnitude range.

Some targets are highly obscured by interstellar dust 
and present a good light curve in the $RP$ band, but very poor in the blue band.
Therefore, amplitudes in the $BP$ band were computed only if the 
average $BP$ magnitudes were brighter than 18.5 mag; 
for the $G$ band a magnitude limit of 17.5 mag was used.
The poor-quality $G$ band curves  of  stars Gaia DR3 5935061172876722176,
Gaia DR3 5972944571113006720, and Gaia DR3 4156489445780902016 
were not considered.
The $BP$-band light curve of Gaia DR3 4256472156970283904 was also 
of poor quality and was discarded.
 
\begin{figure*}[!]
\begin{center}
\resizebox{0.33\hsize}{!}{\includegraphics[angle=0]{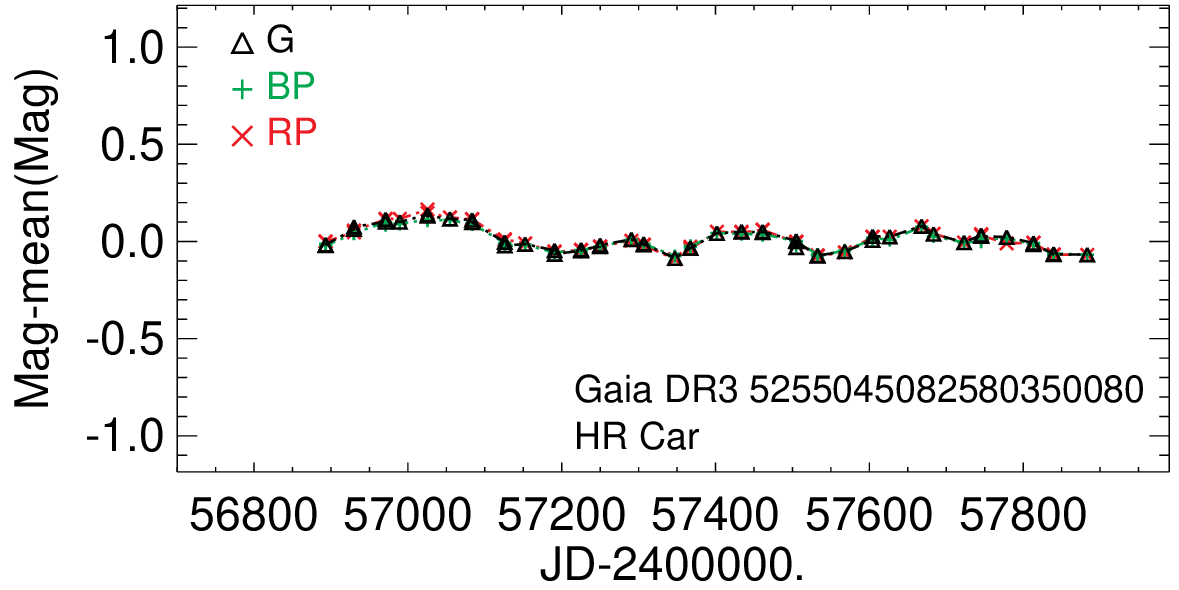}}
\resizebox{0.33\hsize}{!}{\includegraphics[angle=0]{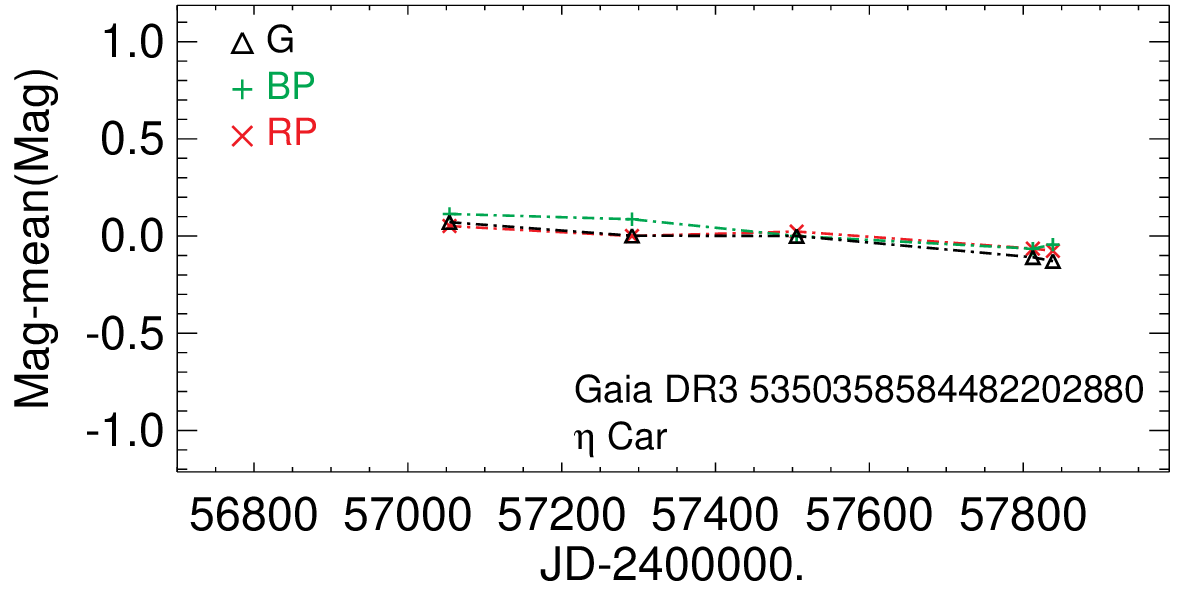}}
\resizebox{0.33\hsize}{!}{\includegraphics[angle=0]{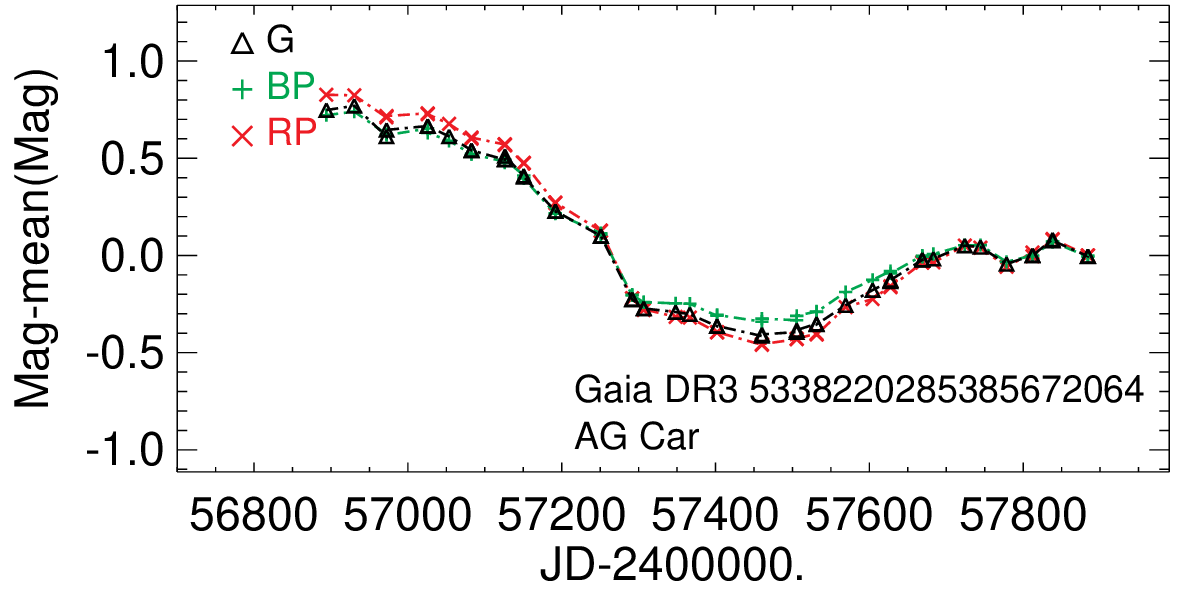}}
\end{center}
\begin{center}
\resizebox{0.33\hsize}{!}{\includegraphics[angle=0]{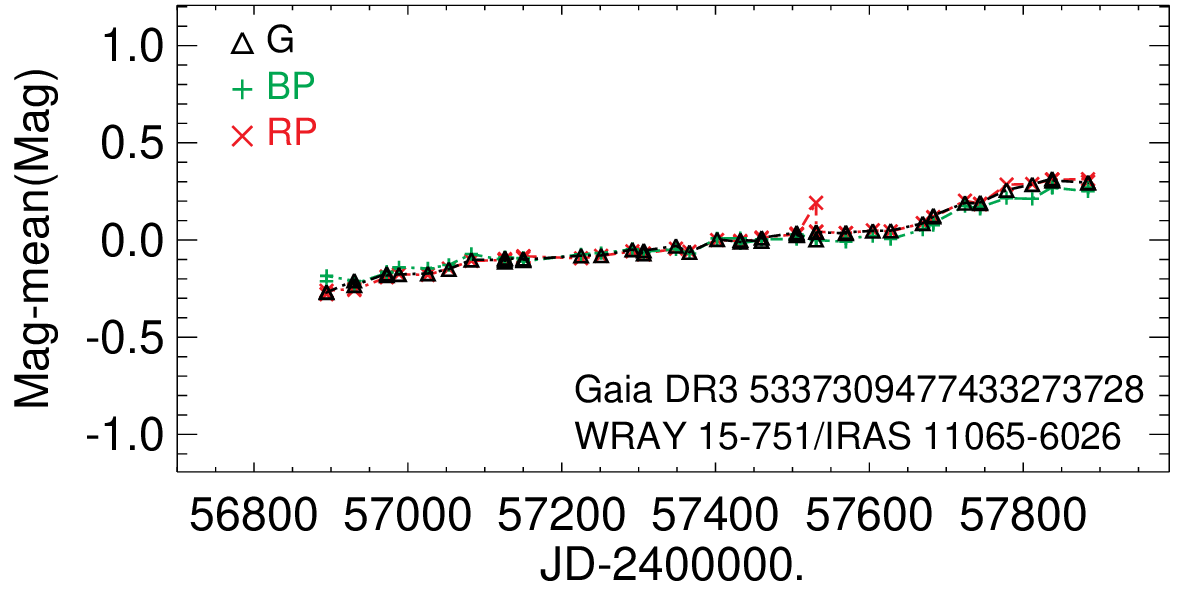}}
\resizebox{0.33\hsize}{!}{\includegraphics[angle=0]{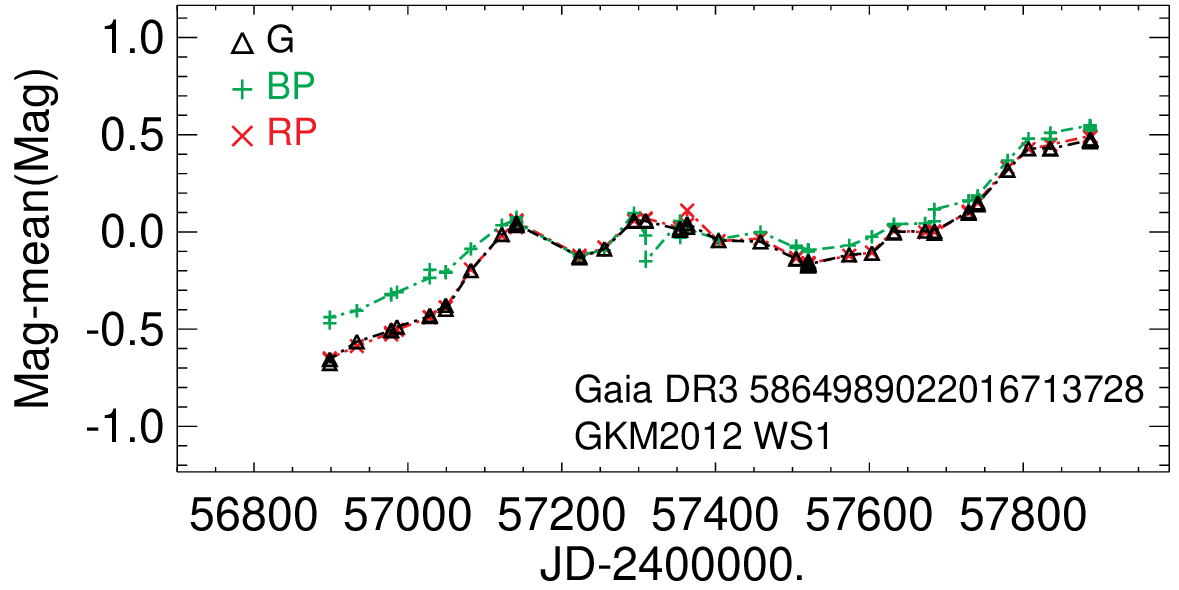}}
\resizebox{0.33\hsize}{!}{\includegraphics[angle=0]{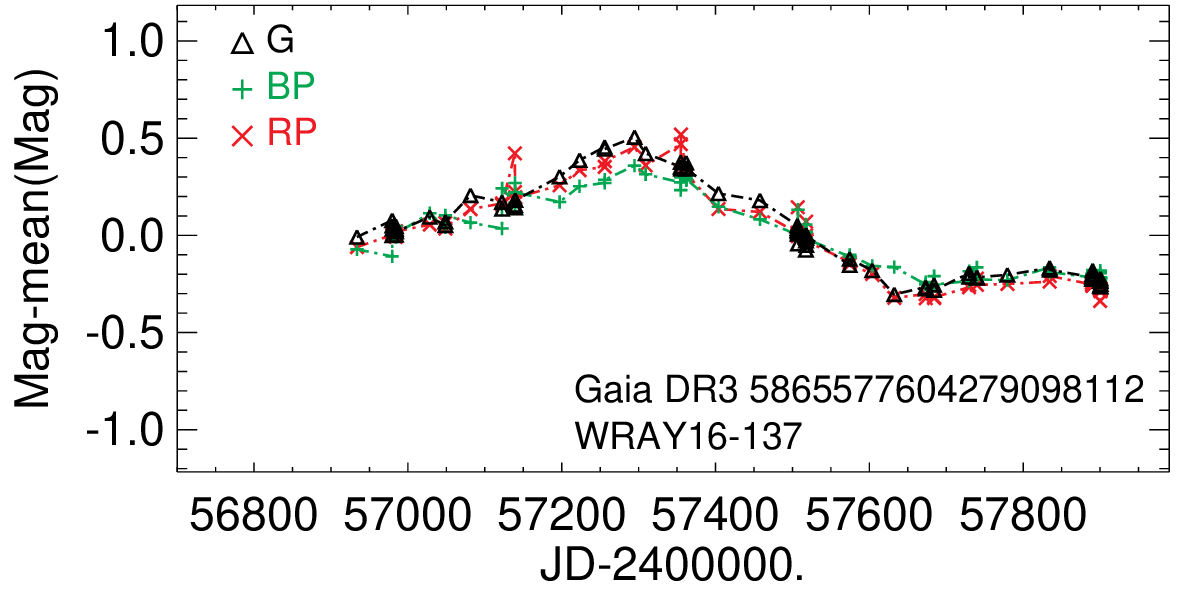}}
\end{center}
\begin{center}
\resizebox{0.33\hsize}{!}{\includegraphics[angle=0]{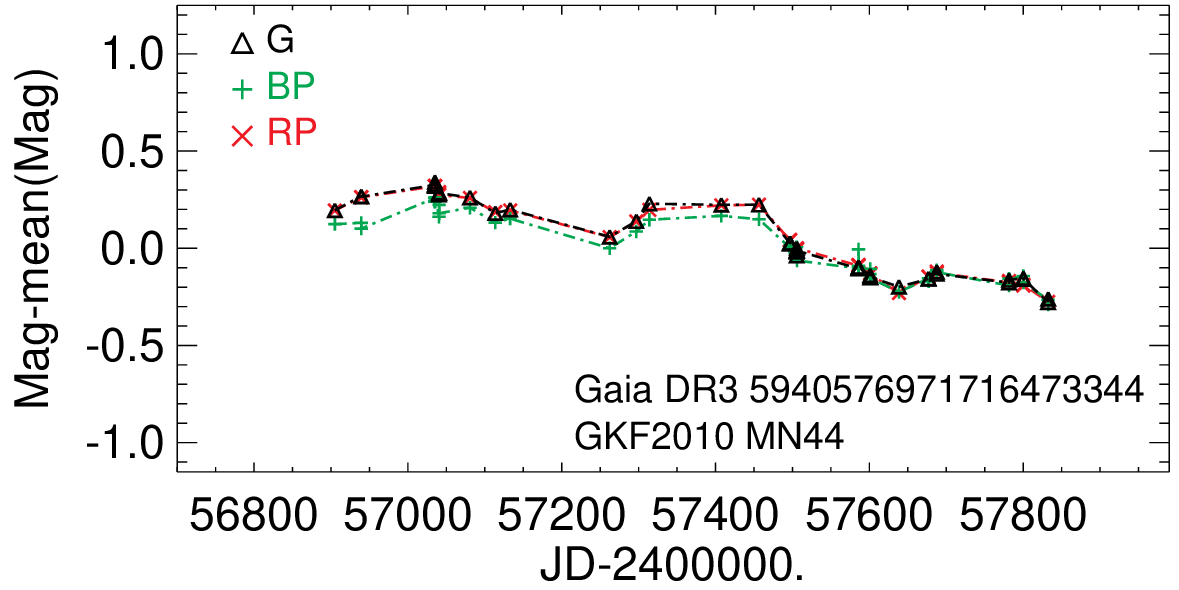}}
\resizebox{0.33\hsize}{!}{\includegraphics[angle=0]{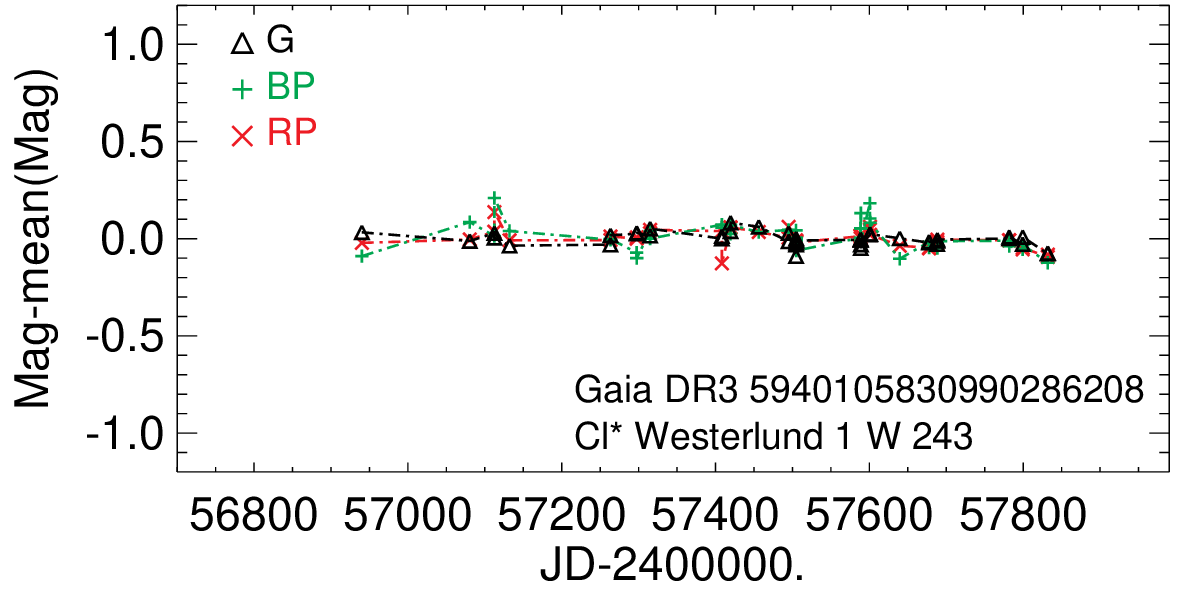}} 
\resizebox{0.33\hsize}{!}{\includegraphics[angle=0]{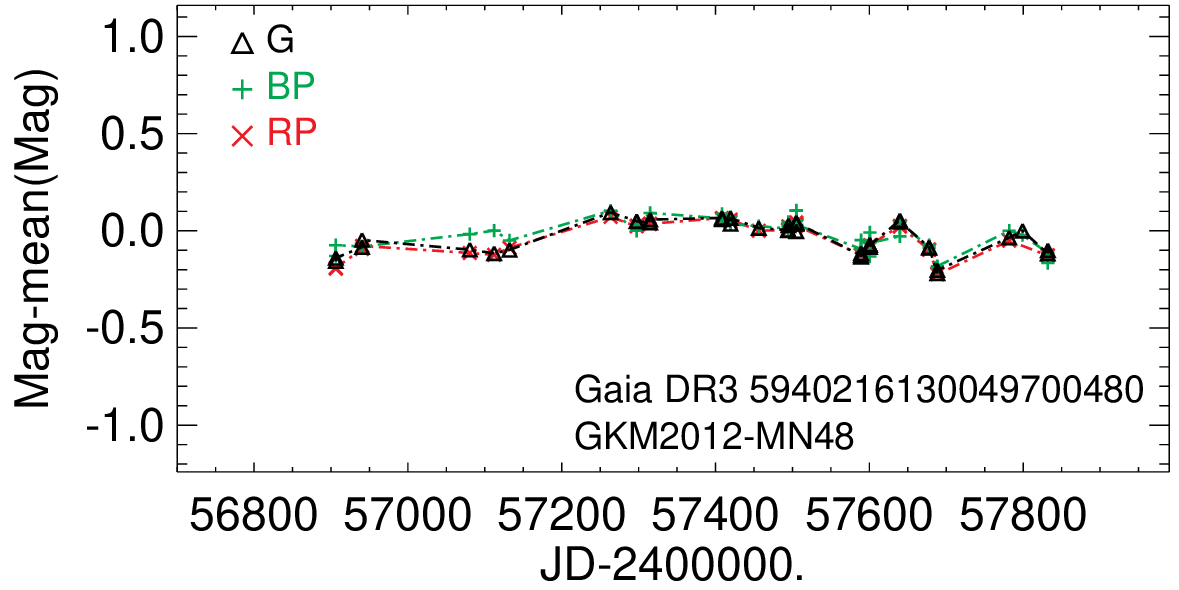}}
\end{center}
\begin{center}
\resizebox{0.33\hsize}{!}{\includegraphics[angle=0]{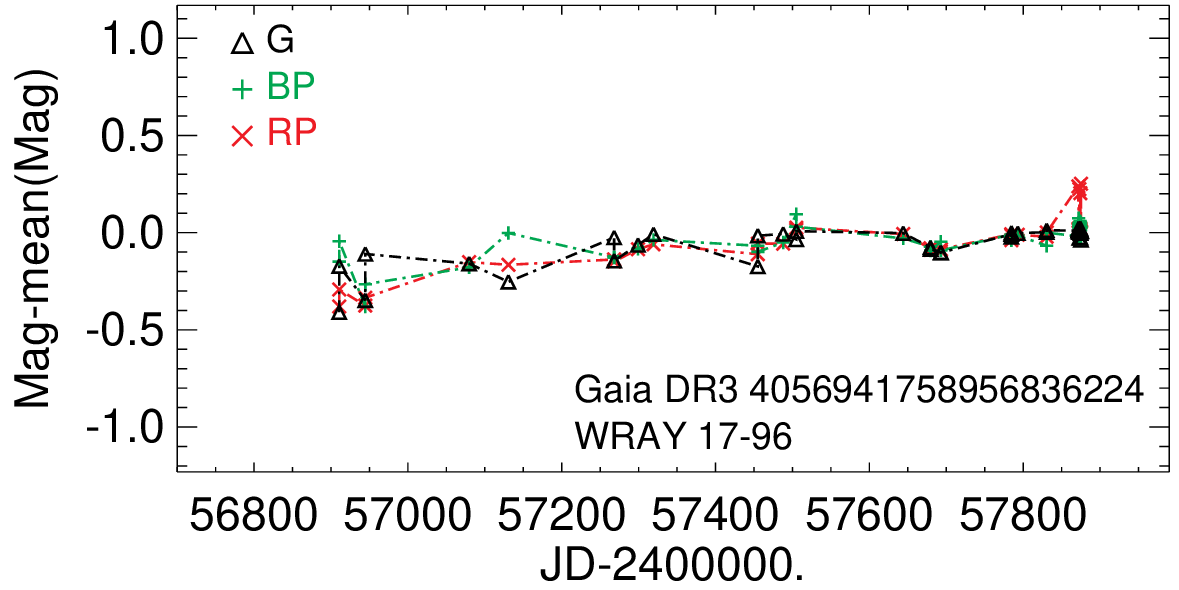}} 
\resizebox{0.33\hsize}{!}{\includegraphics[angle=0]{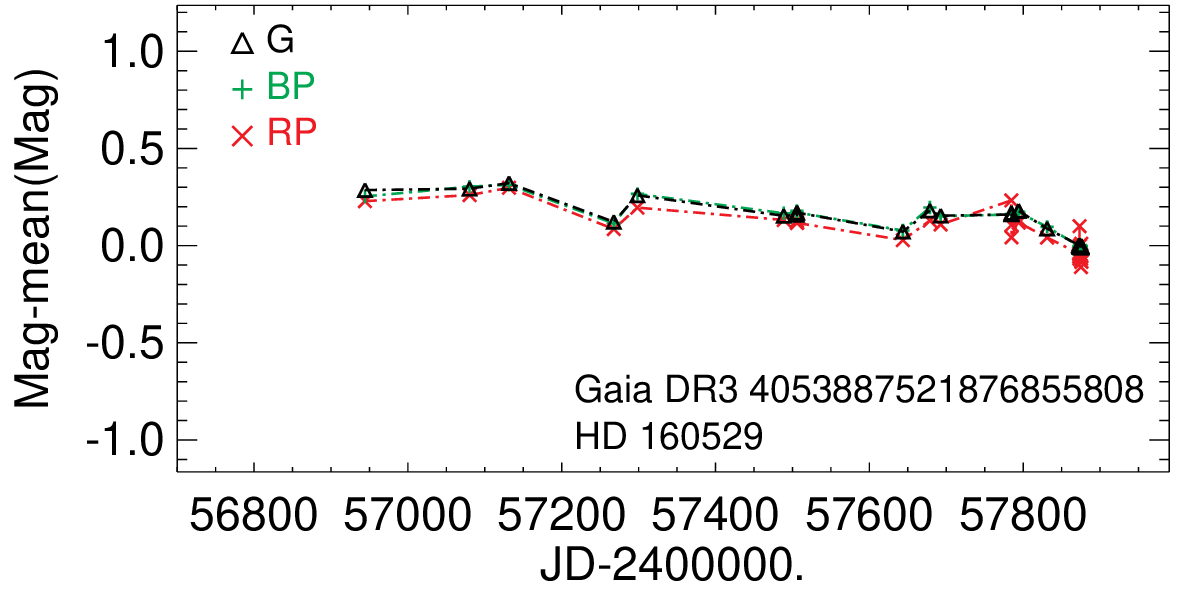}} 
\resizebox{0.33\hsize}{!}{\includegraphics[angle=0]{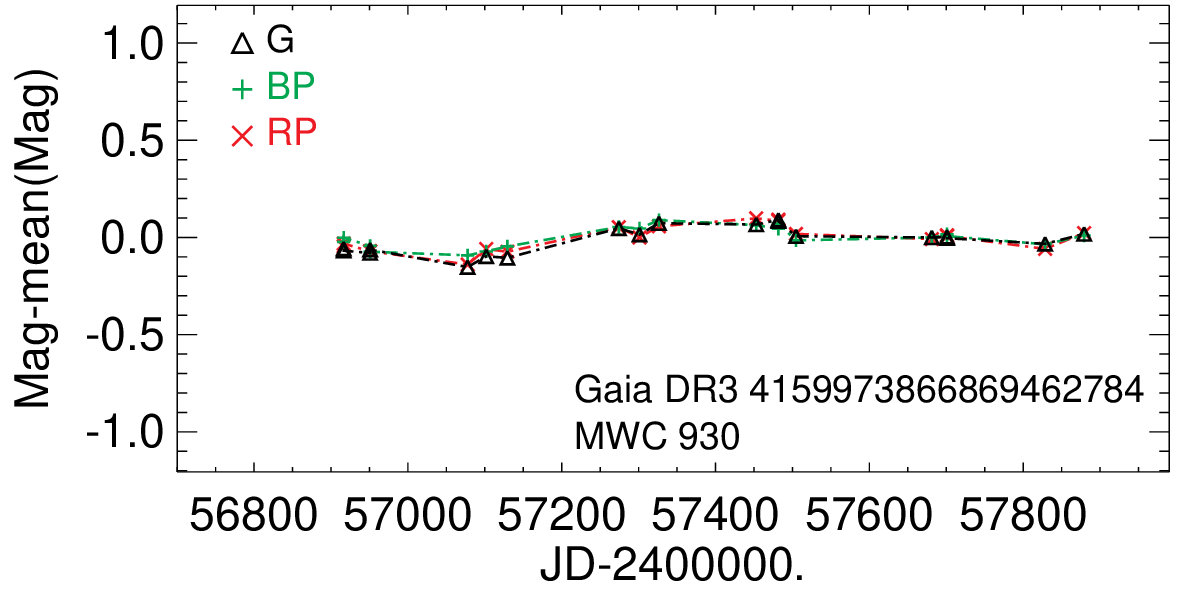}} 
\end{center}
\begin{centering}
\resizebox{0.34\hsize}{!}{\includegraphics[angle=0]{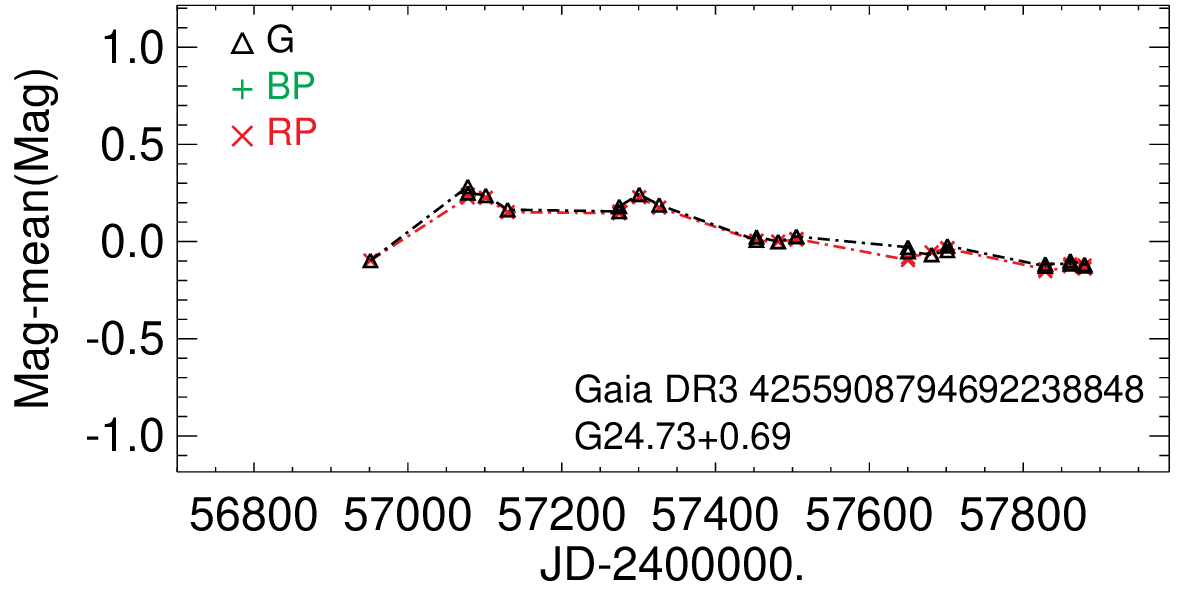}} 
\resizebox{0.34\hsize}{!}{\includegraphics[angle=0]{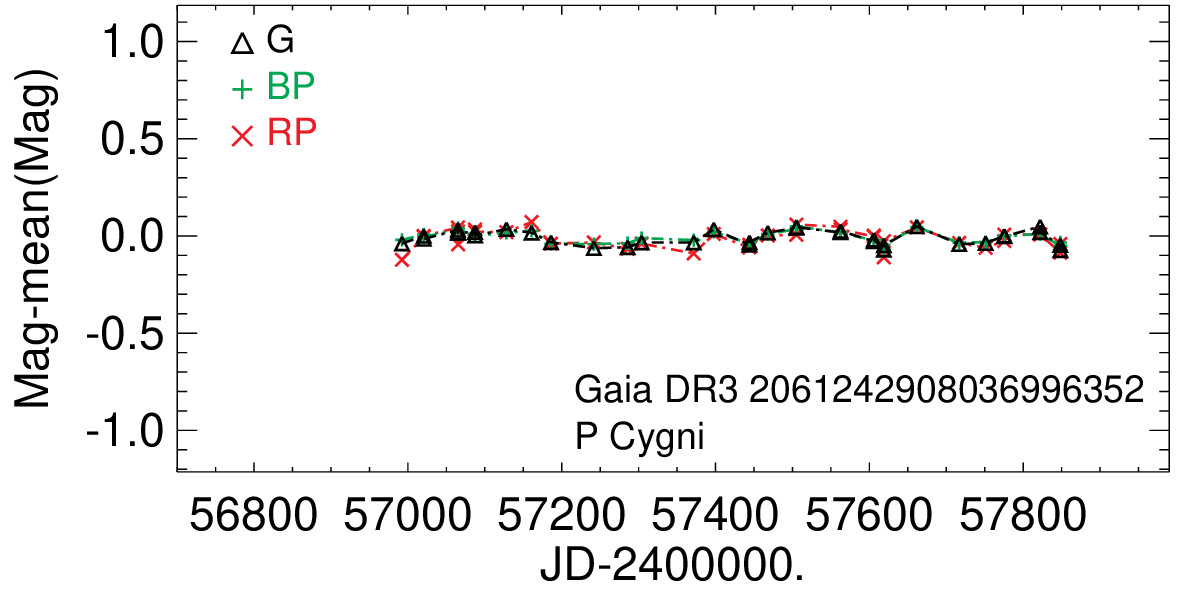}}
\end{centering}
\caption{\label{gaiaLCfig_lbv} Gaia DR3 light curves of known LBVs.}
\end{figure*}

In order to  establish that a star is an LBV, we should witness
a burst of light or show the existence of an S\,Doradus cycle.
The bursts are very sporadic and require decades of observations.
The S\,Doradus cycle operates on timescales of years to decades, 
making it undetectable within the 2.8-year baseline of Gaia DR3. 
Furthermore, Gaia's sparse observing cadence is not well suited 
to capturing the oscillations or microvariations characteristic of LBVs. 
In Gaia DR3 data, 
a notable decrease  of 1.0 mag in less than 400 days 
is observed only in the light curve of the LBV  AG\,Carinae, shown in Fig. 
\ref{gaiaLCfig_lbv},
while WRAY\,16-137 decreases its light of 0.8 mag in 348 days.
Over the Gaia DR3 observed time interval (about 1000 days),
the light curves of the   bona fide LBVs  
 WRAY\,15-751, WS1, MN44,  WRAY\,17-96, and WRAY\,16-137\footnote{ These stars 
are discussed in the literature specifically in terms of 
S Dor-like photometric behavior (WRAY\,15-751 and WRAY\,17-96) 
or S Dor-like outbursts (WS1, MN44, and WRAY\,16-137). 
Gvaramadze's group listed all these stars (but WRAY\,17-96)  
among the bona fide LBVs because of their large observed variability. 
Previously reported $\Delta$mag values are 0.6, 1.45, 1.9, 1.6, and 1.84 mag 
for WRAY\,15-751, WRAY\,17-96, WS1, MN44, and WRAY\,16-137, 
respectively \citep{vangenderen92,buemi11,kniazev15,gvaramadze15,gvaramadze14}.}
show notable brightness changes with amplitudes of 
 0.59, 1.15,  0.60, 0.63, 0.80 mag in the $RP$ band, respectively;
the three bands appear to vary with similar patterns and amplitudes
and their $G-RP$ colors remain largely unchanged — 
BP magnitudes being generally less reliable than $G$-band measurements. 
Clear color-magnitude trends emerge in two cases: 
AG\,Carinae  grows bluer as it dims, while the giant eruptive star  $\eta$\,Carinae  
reddens during its dimming phases. The average $RP$ amplitude of the 14 known LBVs is 0.52 mag 
with a standard deviation $\sigma= 0.35$ mag; the median value is 0.40 mag.
The average variation in the $G$ band is 0.48 mag with $\sigma= 0.34$ mag,
and in the $BP$ band is 0.45 mag with $\sigma= 0.30$ mag.
It is not possible to firmly reidentify LBVs solely on the basis of
Gaia DR3 light curves. Beyond variability studies, 
the new high-quality light curves also prove valuable for identifying 
and characterizing binary systems.

\begin{figure*}[!]
\begin{center}
\resizebox{0.33\hsize}{!}{\includegraphics[angle=0]{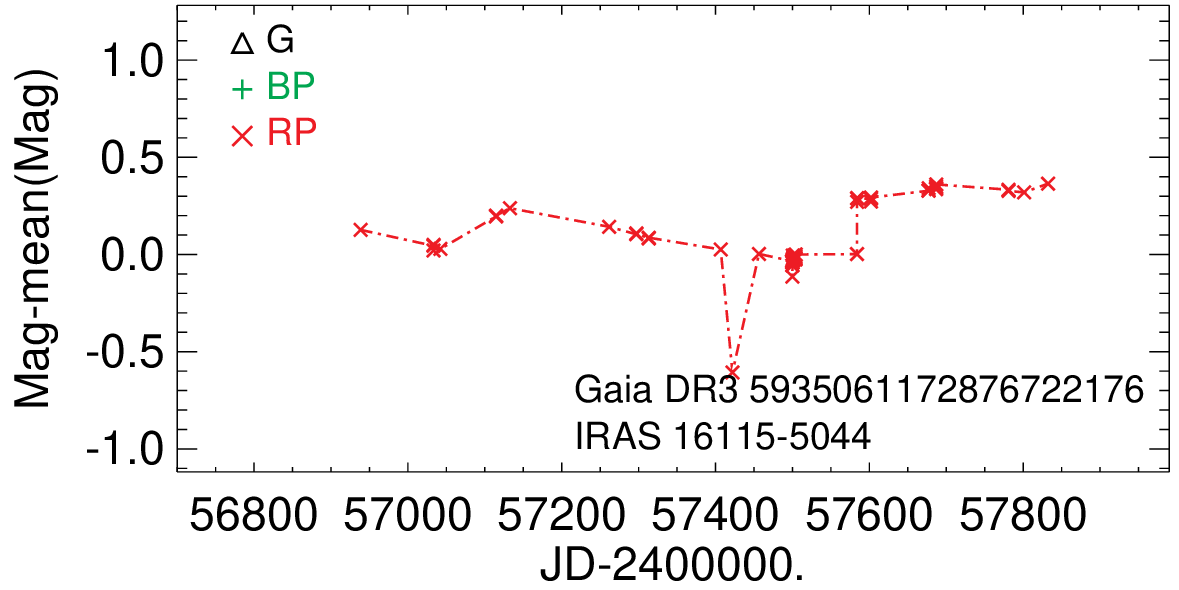}} 
\resizebox{0.33\hsize}{!}{\includegraphics[angle=0]{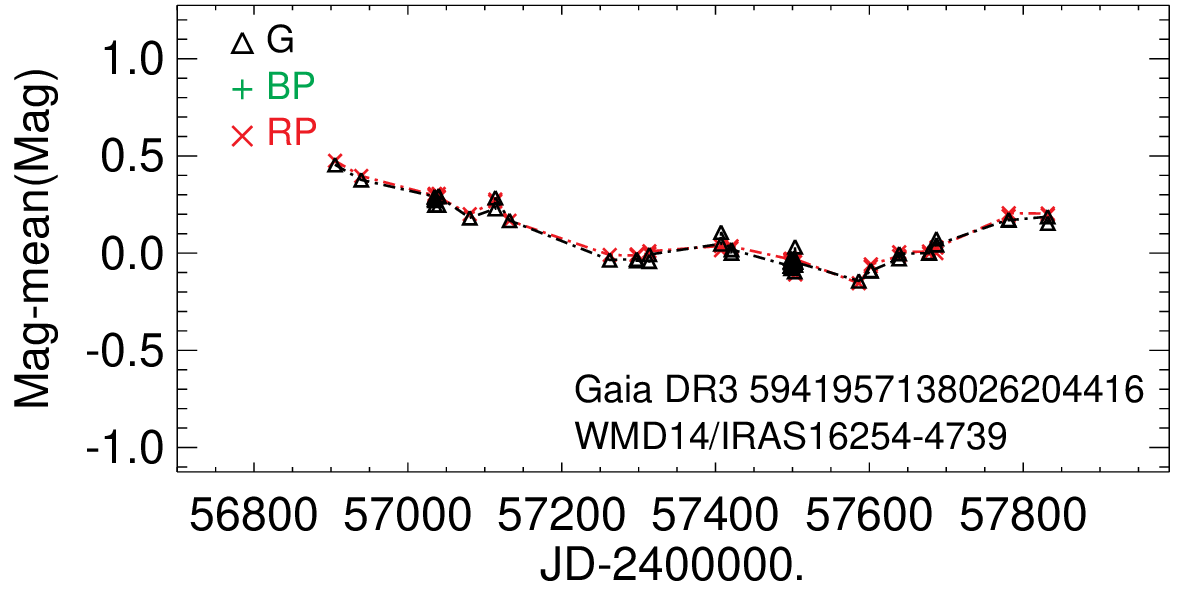}}
\resizebox{0.33\hsize}{!}{\includegraphics[angle=0]{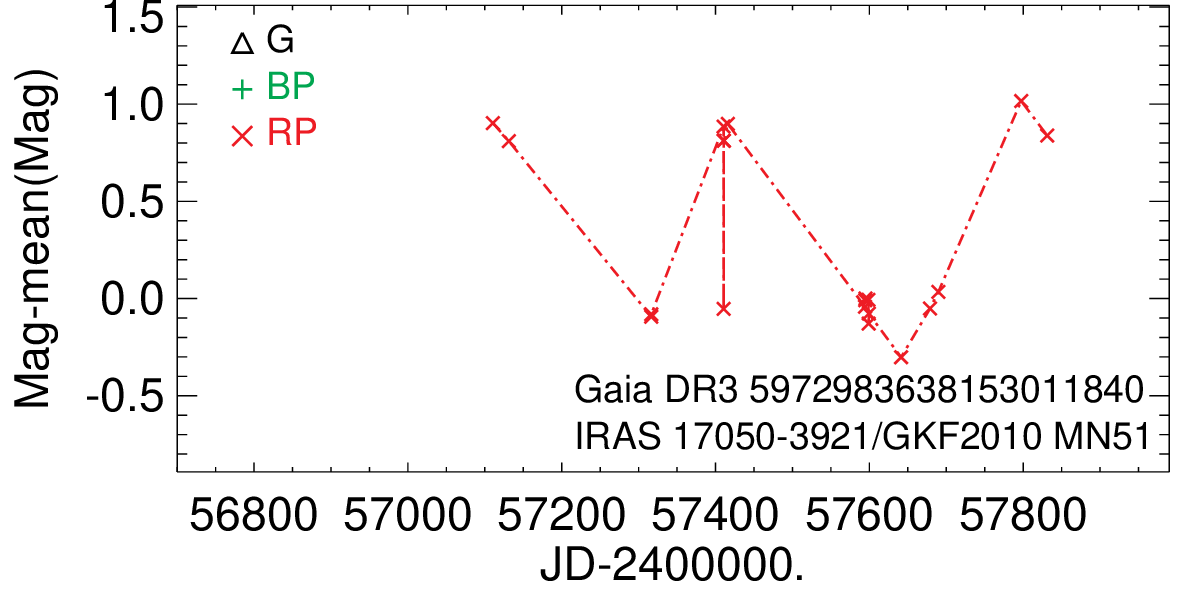}}
\end{center}
\begin{centering}
\resizebox{0.33\hsize}{!}{\includegraphics[angle=0]{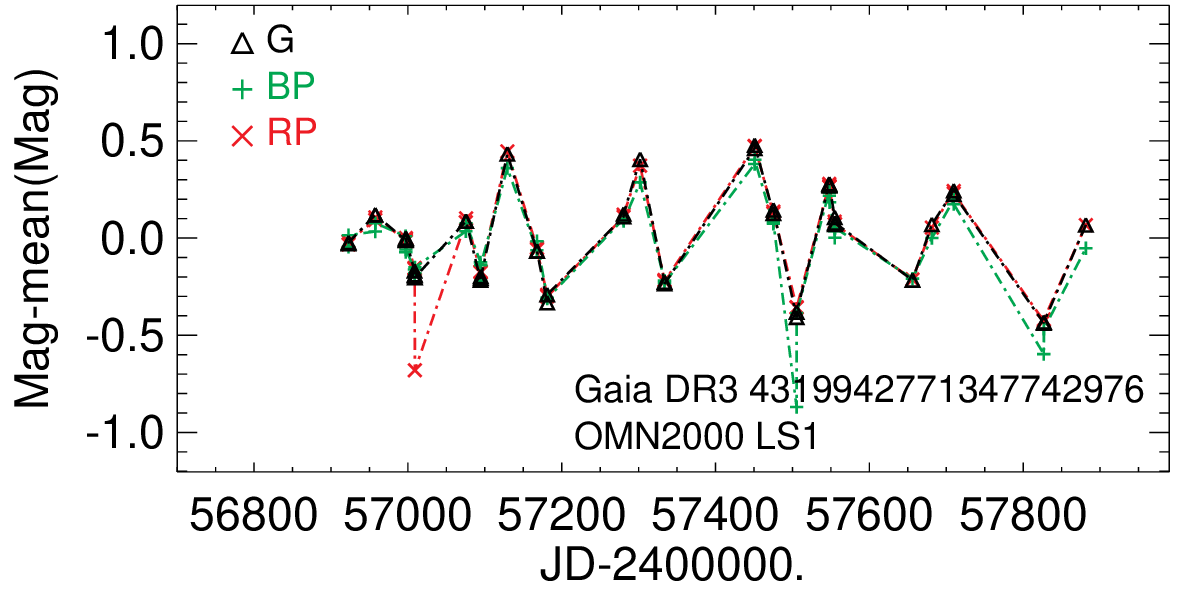}}
\resizebox{0.33\hsize}{!}{\includegraphics[angle=0]{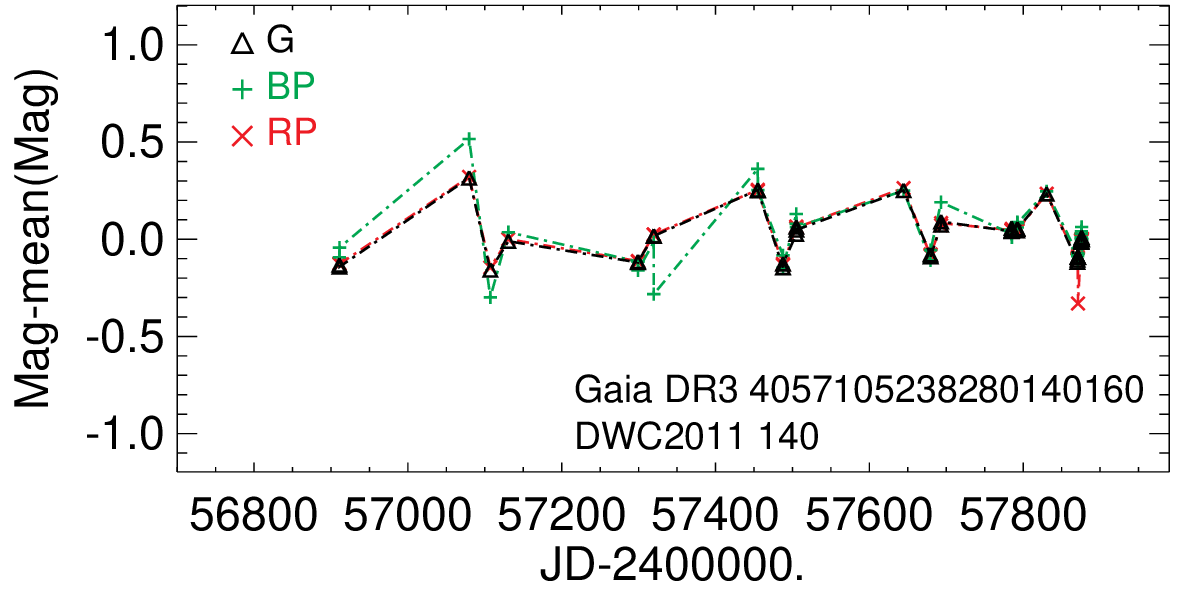}}
\end{centering}
\caption{\label{gaiaLCfig_clbv} Gaia DR3 light curves of 
cLBVs with detected magnitude variations larger than 0.5 mag.}
\end{figure*}

Therefore, we use 0.6 mag as a threshold to select cLBVs for further 
inspection of their light curves but  this value is somewhat 
arbitrary and it should only be used 
to motivate further investigation using other techniques, 
not to firmly classify a star as an LBV.
Large variations of 0.97,  0.63, 1.32,  1.15, 0.65 mag 
are detected   in the $RP$ light curves of  five cLBVs 
(IRAS\,16115$-$5044, WMD2010\,14, GKF2010\,MN51,  
OMN2000\,LS1, and DWC2011\,140); see Fig. \ref{gaiaLCfig_clbv}.

\subsection{The Bochum survey}
The Bochum survey provides $r'$ and $i'$ 
time-series photometry for  southern variables  (10--15 mag)
over a three-year baseline \citep{haas12},
using a camera with a pixel scale of 2\farcs4.
Its value here is not only the additional amplitude 
information, but also the denser temporal coverage relative 
to Gaia for some southern objects, which helps distinguish 
monotonic or slowly varying LBV-like behavior 
from periodic variability.

\begin{figure*}[!]
\sidecaption
\resizebox{0.33\hsize}{!}{\includegraphics[angle=0]{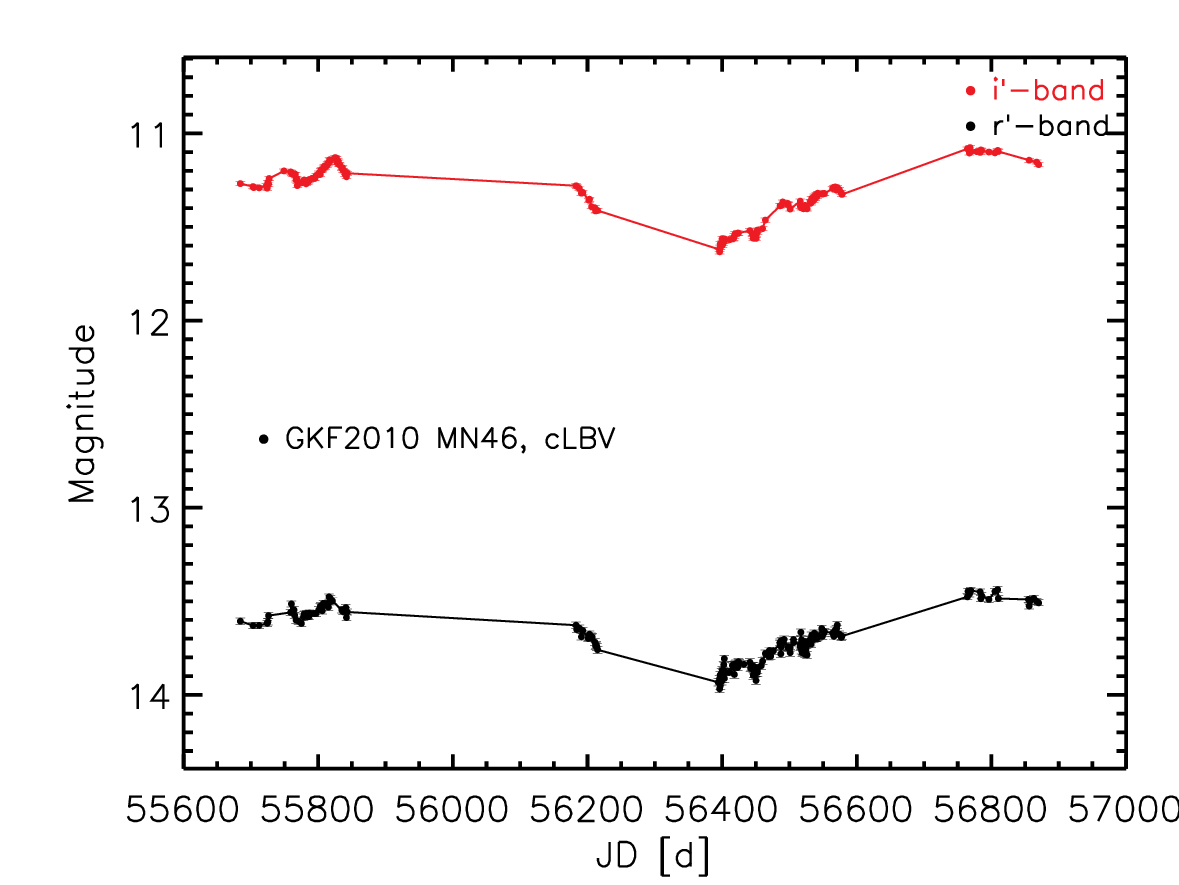}}
\resizebox{0.33\hsize}{!}{\includegraphics[angle=0]{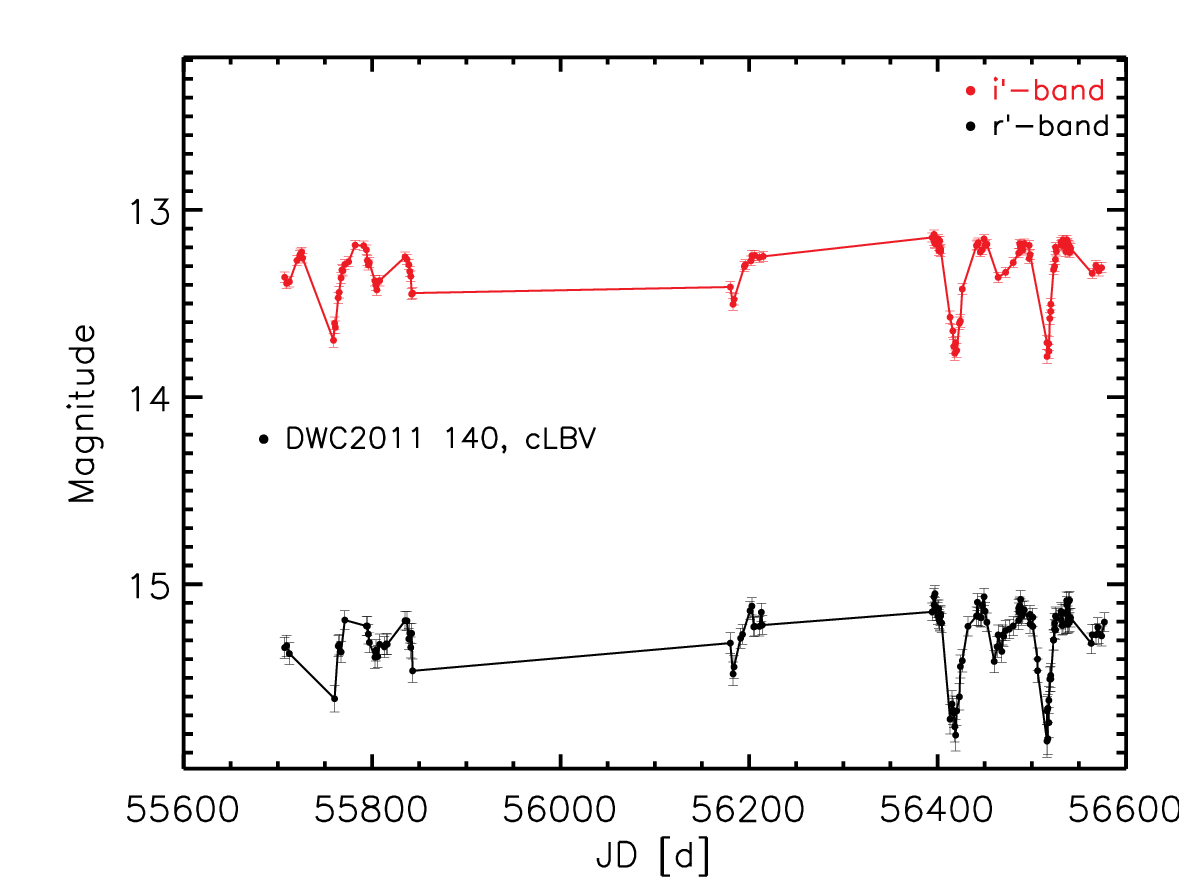}}
\caption{\label{bochumLCfig} Light curves from the Bochum survey of 
detected cLBVs with magnitude variations larger than 0.5 mag.}
\end{figure*}

We found data in the Bochum survey for eight LBVs and six cLBVs.
All but two known LBVs have magnitude variations larger than 0.6 mag.
 HR\,Carinae  and W243 were found in a quiet phase and 
varied only 0.2--0.25 mag. 
The candidates  GKF2010\,MN46, and DWC2011\,140 
show variations of $\approx 0.6$ mag in both $r'$ and $i'$ mag;
see Fig. \ref{bochumLCfig}.

\subsection{Additional curves used to characterize  
the Gaia and/or Bochum large variables}

ASAS\,SN and ZTF data
were used to test whether the large-amplitude 
Gaia and Bochum variables show coherent periodicity over longer 
or better-sampled baselines. ASAS\,SN supplies multiyear $V$-band 
light curves of bright sources  ($11 < V < 17$ mag), 
with a nightly cadence and  a  spatial resolution of $\approx 16$\arcsec\ 
\citep{jayasinghe19},
while ZTF provides higher-cadence (once or three times per night)
optical photometry (in bands $g$ and $r$) at better 
angular resolution (2\farcs0) 
in the northern sky \citep{chen20}. 
These complementary data are especially important for recognizing 
eclipsing-binary-like light curves among cLBVs.

Twenty-four of the 75 LBV and cLBV stars are included 
in Jayasinghe’s work. 
Fifteen are classified as SDOR, three as SR (semi-regular),
and six as L (long) variables. 
The estimated variations (minimum magnitudes - maximum magnitudes)
and  ASAS\,SN classification are listed in 
the Appendix \ref{amplitudes.table}.\\
Among the  cLBVs with large variations detected in Gaia,
only the OMN2000\,LS1 curve was available 
in the ZTF survey, 
as shown in Sect. \ref{variability.section}. 

\section{ Results: Detected large variations in cLBV}
\label{variability.section}

Gaia DR3 discovers variations $\ga 0.6$ mag
in the light curves of five  cLBVs, while the Bochum survey adds two more 
cLBVs (with one in common with the detected Gaia).

\subsection{Nonperiodic light curves}

The light curves of IRAS\,16115$-$5044, WMD2010\,14, and MN46
appear nonperiodic and consistent with those of other known LBVs 
-- slowly rising and/or decreasing.\\
The analysis of $K$-band spectra of IRAS\,16115$-$5044 (``the bullet'') 
is presented and modeled by \citet{figer20}. 
The richness of the spectrum displaying 
\ion{He}{i}, \ion{H}{i}, \ion{Fe}{ii}, 
[\ion{Fe}{ii}], \ion{Mg}{ii}, \ion{Na}{i}  lines in emission, the P\,Cygni profiles,
and the  variability of the line strengths ensured an LBV, even though
confirmation from a photometric light curve was missing. 

WMD2010\,14 (IRAS\,16254$-$4739, 2MASS J16290377$-$4746264)  
was spectroscopically observed in the  $JHK$ band by \citet{wachter10}.
 It is  featured with \ion{He}{i}, \ion{H}{i}, \ion{Mg}{ii}, 
\ion{Fe}{ii}, and [\ion{Fe}{ ii}] lines in emission.

An optical spectrum of the nebulae around GKF2010\,MN46 
with  numerous \ion{Fe}{ii} and [\ion{Fe}{ii}] lines, typical of LBVs, 
is presented by \citet{gvaramadze10}. 
The spectrum is reanalyzed by \citet{arun25};
in addition to identifying more than 100 lines with P\,Cygni profiles, 
the spectra show also singly ionized lines from \ion{N}{ii}, \ion{Si}{ii}, 
\ion{Al}{ii}, \ion{Ca}{ii}, and \ion{Mg}{ii}, and forbidden [\ion{N}{ii}] lines.
The authors suggest a \Teff\ from 12,000 to 30,000 K.
Interferometric observations of 
GKF2010\,MN46 confirm a binary with  a companion (or disk feature)  
3.3 mas away and 3.7 magnitude fainter  \citep{mahy22}.

\begin{table*}
\caption{ \label{table:periods} Periods and periodic variables.}
\begin{tabular}{l|rrr|rrr}
\hline
\hline
Name             &  \multicolumn{3}{c}{This work - Period} & \multicolumn{3}{c}{Literature Period}\\
                 &Gaia       & ASAS\,SN     & ZTF    &  Gaia  & ASAS\,SN   & ZTF \\
                 &LS/PDM     & LS/PDM      & LS/PDM &    &    &  \\
                 &[d]        & [d]          &  [d]  &    [d] &  [d]   &  [d] \\
\hline
GKF2010\,MN51      & 390.29/388.08? & 27.35/27.68?&              &           &              &    \\
OMN2000\,LS1 &  79.86/80.82   & 80.85/80.27 &  81.91/80.29 &80.48      &  80.63       & 79.47   \\
DWC2011\,140       & 95.07/94.85   & 46.42/95.07 &              &0.015      &  47.74       &    \\
\hline
\end{tabular}
\begin{list}{}
\item For each available light curve, the LS and PDM periods are measured. 
Question marks identify uncertain periods for which the Gaia and ASAS\,SN 
solutions are inconsistent.
\end{list}
\end{table*}

\subsection{Periodic light curves}
Three cLBVs show apparent periodicity in Gaia DR3: 
GKF2010\,MN51, 
OMN2000\,LS1, and DWC2011\,140.\\ 
GKF2010\,MN51 (WMD2010\,23a, IRAS\,17050$-$3921, 2MASS\,J17082913$-$3925076)   
was spectroscopically observed in the $JHK$ band by \citet{wachter10}.
Its spectrum shows features due to the emission lines \ion{He}{i}, \ion{H}{i}, \ion{Mg}{ii}, 
\ion{Fe}{ii},  \ion{Na}{i}, and [\ion{Fe}{ ii}].
  
A $K$-band spectrum of OMN2000\,LS1
is shown and modeled by  \citet{clark09b}.
The spectrum is dominated by a strong \ion{He}{i} line at 2.05 \um\ in emission
with a P\,Cygni profile and a strong \brgamma\ emission.
\ion{Mg}{ii} lines (2.138 \um\ and 2.144 \um) are detected, but not \ion{Fe}{ii} lines. 

DWC2011\,140 has been observed spectroscopically at least twice  
\citep{clark21}.
Indeed, \citet[their star n. 1,][]{mauerhan10b} established a B0--2 I type 
based on a  clear \ion{He}{i} at 2.05 \um\ in emission with a blue absorption wing, 
strong \brgamma\ emission, and a line emission \ion{Na}{i}.
The other $K$-band spectrum by \citet{geballe19} shows lines 
from \ion{Fe}{ii}, \ion{Mg}{ii}, \ion{Na}{i}, \ion{He}{i}, 
a \brgamma\ emission line with a  blue-wing  
absorption, and CO bandheads,   which are a typical signature 
of windy B[e]SG stars \citep{kraus19}.  However, its GLIMPSE
colors are not as red as those of known B[e] stars 
(see discussion in Sec.\ \ref{sect.periodic}).

The periods for OMN2000\,LS1 and DWC2011\,140 are robust because 
independent surveys and PDM 
analyses recover consistent values. 
The period of GKF2010\,MN51 remains uncertain 
because the Gaia and ASAS\,SN solutions disagree and it is therefore 
not used as strong evidence for binarity.

The three  Gaia DR3 light curves were analyzed
with an LS periodogram\footnote{The python code 
 is distributed in astropy.timeseries.
The minimum P is set to 0.1 d and maximum P
is set as  $2 \times {\rm (max - min)}$ d (1700-1900 d).} 
\citep[][]{vanderplas18}.
The frequencies with the highest power are retained,
which correspond to  P = 390.29, 79.86, and 95.07 d, 
respectively, as listed in Table \ref{table:periods}.
The false alarm probability is
FAP=2.00e-16 for MN51, FAP=9.79e-15 for LS1, 
FAP=2.02e-11 for DWC2011\,140.

The periods were confirmed using PDM\footnote{The python code   
follows \citet[][]{stellingwerf78}.
For the period search of OMN2000\,LS1 and DWC2011\,140
a range of 50 to 150 d was used, while for GKF2010\,MN51
the search period was enlarged to 10-400 d.
Five phase bins were used and it was required that at least 
half of the bins were populated.}
\citep[][]{stellingwerf78}, 
which is better suited than the LS method 
for binary light curves whose shape departs from a pure sinusoid.
The PDM yields more robust periods as it makes no 
assumption about the shape of the light curve.
The best-fit  period values are 388.08, 80.82, and 94.85\,d, respectively.
The comparison of the two sets of periods
ensures that reliable  \textit{Gaia} periods were obtained 
for OMN2000\,LS1 and DWC2011\,140.
For OMN2000\,LS1, our best \textit{Gaia} period (80.82 d) 
agrees with that (P=80.48 d) 
reported in the  \textit{Gaia} DR3 catalog of long period 
variable stars of \citet{lebzelter23}, the  semi-amplitude 
(half of the peak-to-peak amplitude) is 0.39 mag. 

\begin{figure}[!]
\begin{centering}
\resizebox{1\hsize}{!}{\includegraphics[angle=0]{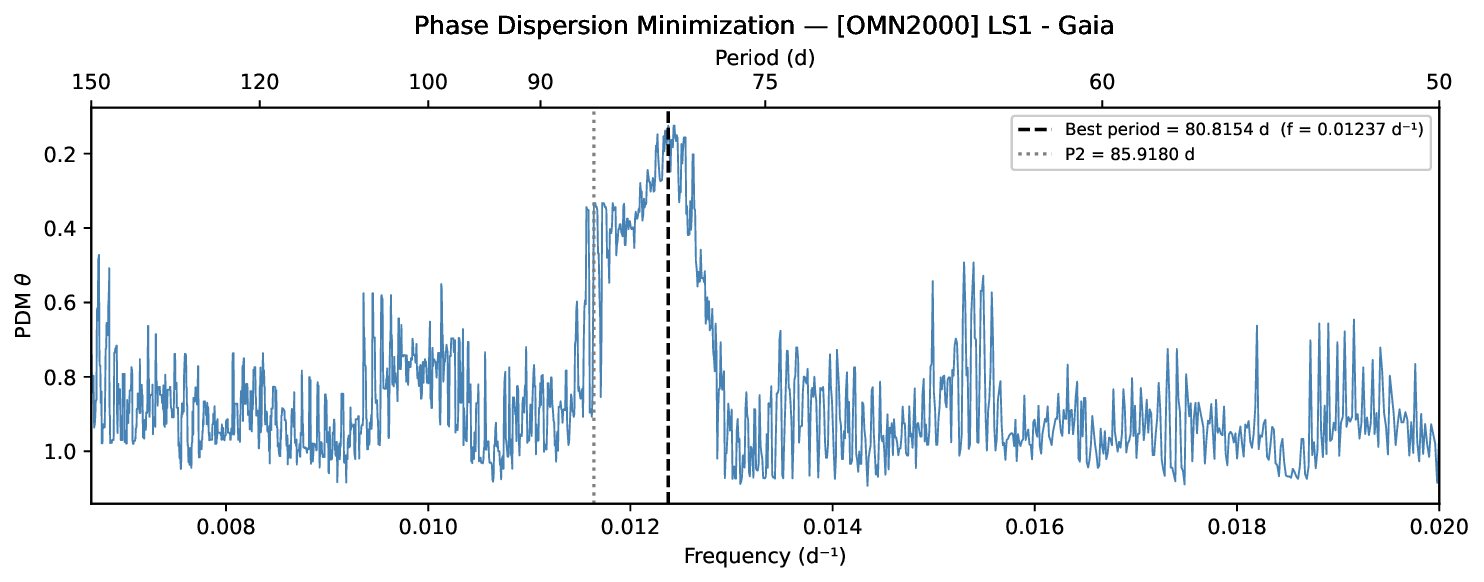}}
\resizebox{1\hsize}{!}{\includegraphics[angle=0]{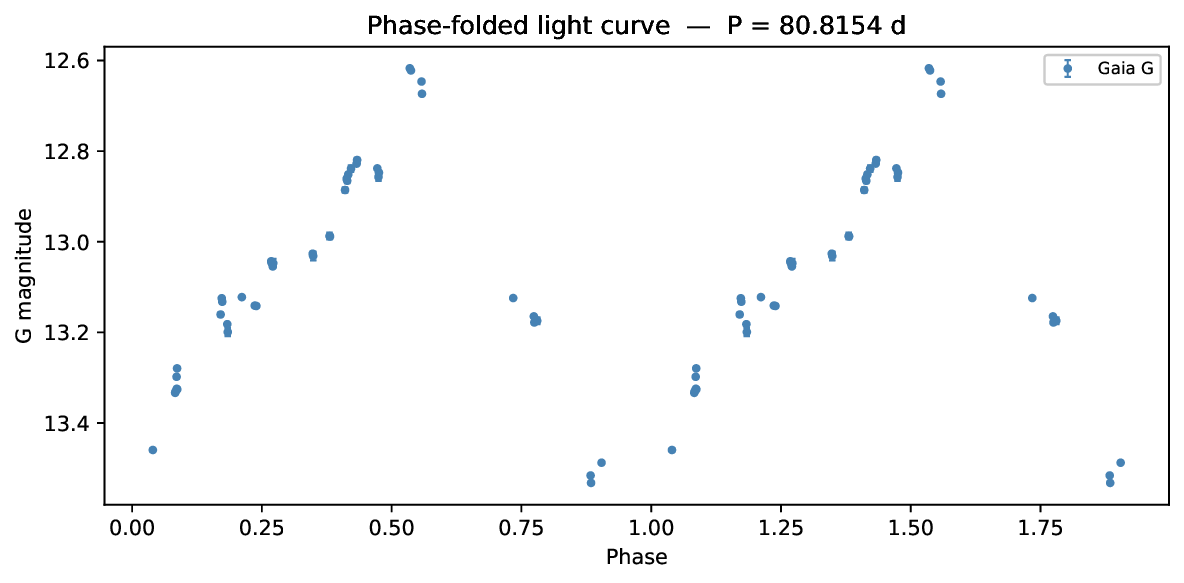}}
\end{centering}
\caption{\label{W51-per} OMN2000\,LS1. {\it Upper panel:}
$G$-band PDM periodogram of the \textit{Gaia} DR3 time series. 
The dashed black line marks the 
highest-probability frequency.
{\it Lower panel:} \textit{Gaia} DR3 phase diagram.
Filled cyan circles mark the $G$-band data points. 
The period adopted is marked on the figure title.
}
\end{figure}

\begin{figure}[!]
\begin{centering}
\resizebox{1\hsize}{!}{\includegraphics[angle=0]{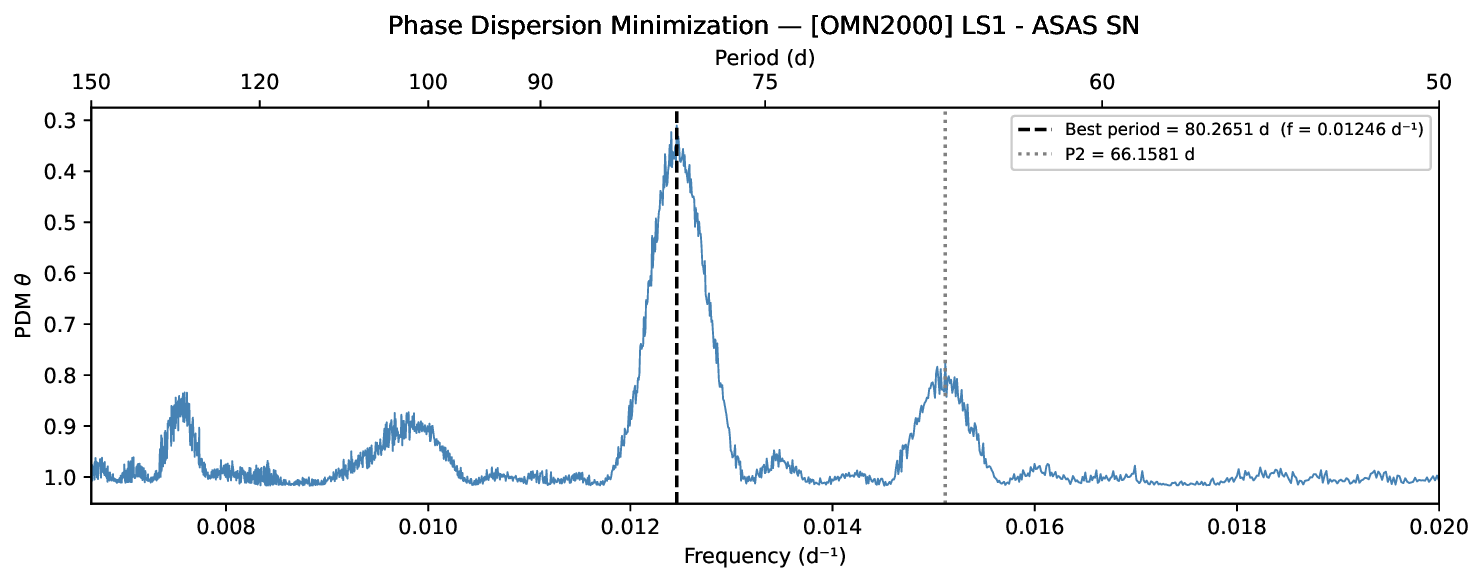}}
\resizebox{1\hsize}{!}{\includegraphics[angle=0]{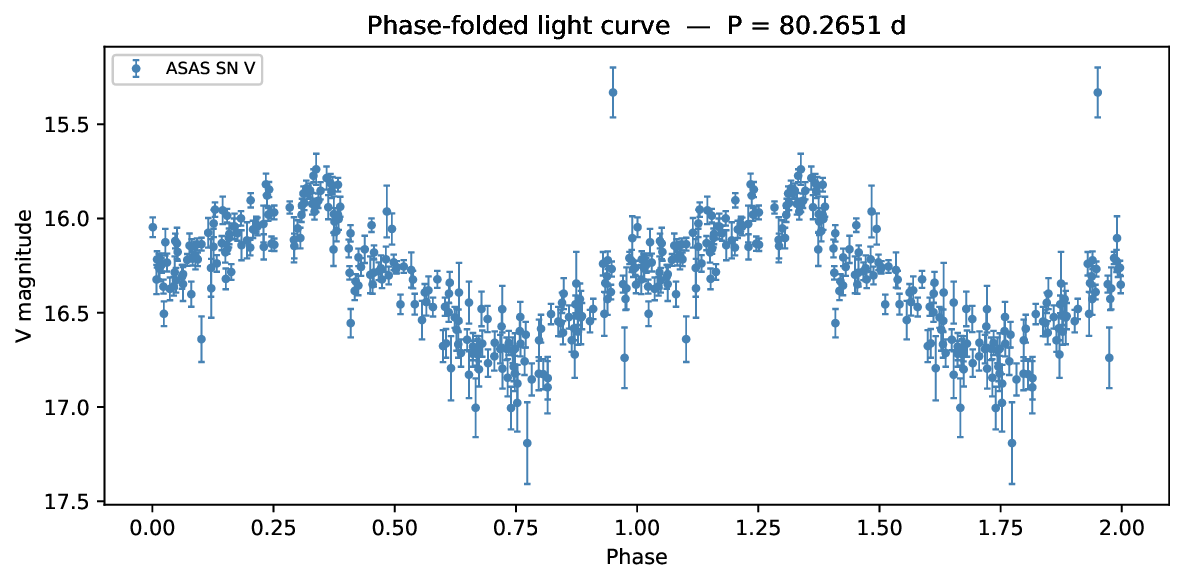}}
\end{centering}
\caption{\label{W51-per2} OMN2000\,LS1 {\it Upper panel:}
$V$-band PDM periodogram of the ASAS\,SN time series. The dashed black line marks the 
highest-probability frequency.
{\it Lower panel:} ASAS\,SN $V$-band phase diagram.
The best ASAS\,SN period is adopted, as  marked.
}
\end{figure}

\begin{figure}[!]
\begin{centering}
\resizebox{1\hsize}{!}{\includegraphics[angle=0]{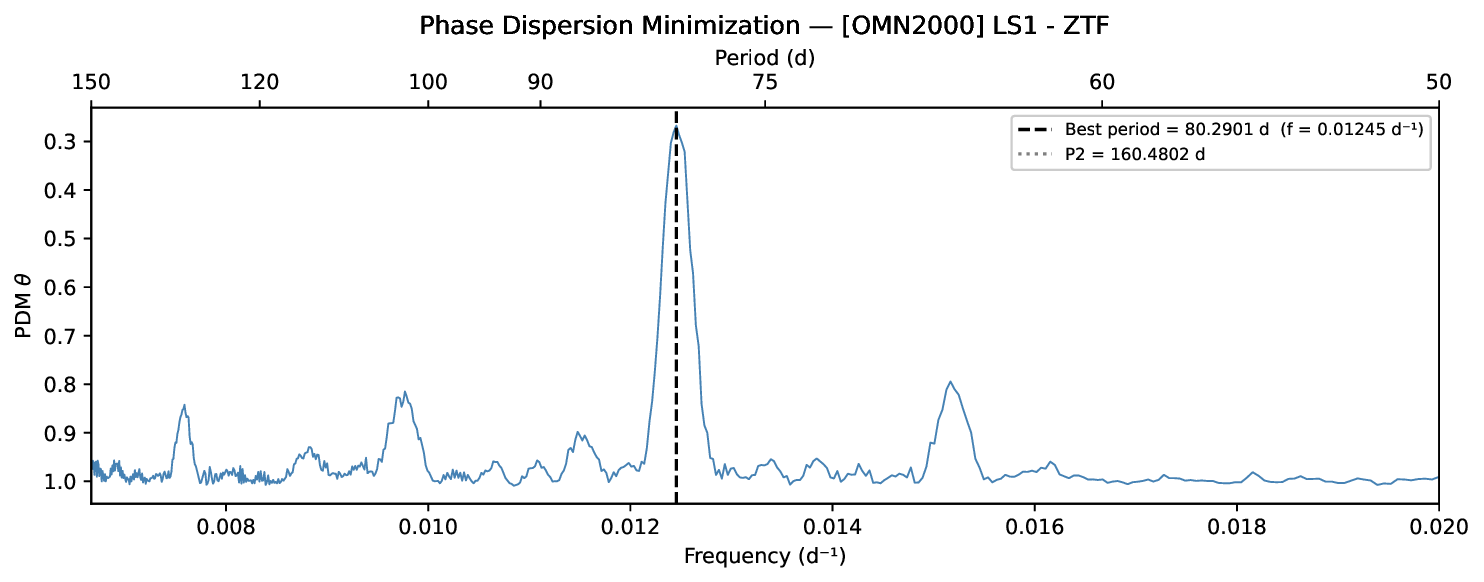}}
\resizebox{1\hsize}{!}{\includegraphics[angle=0]{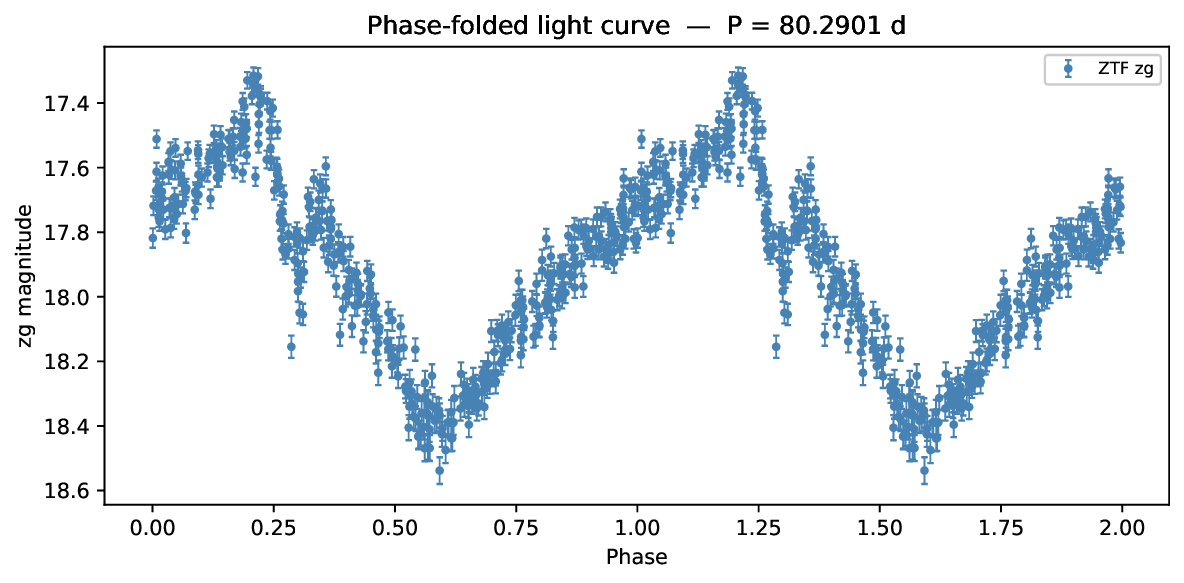}}
\end{centering}
\caption{\label{W51-per3} OMN2000\,LS1.  {\it Upper panel:}
$zg$-band PDM periodogram of the ZTF time series. 
 The dashed black line marks the 
highest-probability frequency  P = 80.29 d, the ZTF catalog lists
P= 79.47 d.
{\it Lower  panel:} ZTF  phase diagram.
Filled cyan circles mark the $zg$-band data points. 
P=80.29 d is adopted, as  marked.
}
\end{figure}

Light curves from ASAS\,SN and ZTF were also analyzed 
when available. The ASAS\,SN has a poor spatial resolution (8\arcsec).
The ZTF with the high cadence  
(minutes-to-hours during visits) running since 2017
and with 1\arcsec\ resolution is ideal to detect 
light curve features and secondary minima \citep{chen20}.
OMN2000\,LS1 was also observed by the ZTF and ASAS\,SN surveys,
and P= 79.47 d  \citep{chen20} and 80.63 d 
\citep[VIZIER II/366/catv2021,][]{jayasinghe19b} are given. 
We obtained P=80.85 d with FAP = 2.859e-54 (ASAS\,SN) 
with the LS method.
For the ZTF curve, P= 1.010 d with FAP = 5.694e-87,
and P= 81.91 d with a lower FAP of 1.193e-12.
With the PDM method we obtained   slightly smaller values,
P=80.27 d (ASAS\,SN) and 80.29 d (ZTF);
moreover, the ASAS\,SN folded light curve presented a clear
drop and/or feature at phase 0.75 and the ZTF folded high-curve, 
which has the fastest cadence, 
revealed a clear secondary minimum (asymmetric). Therefore, we 
retained as the best period the  PDM period from the ZTF curve.
The light curves are shown in Figs. \ref{W51-per}, \ref{W51-per2}, 
and \ref{W51-per3}. 

\begin{figure}[!]
\begin{centering}
\resizebox{1\hsize}{!}{\includegraphics[angle=0]{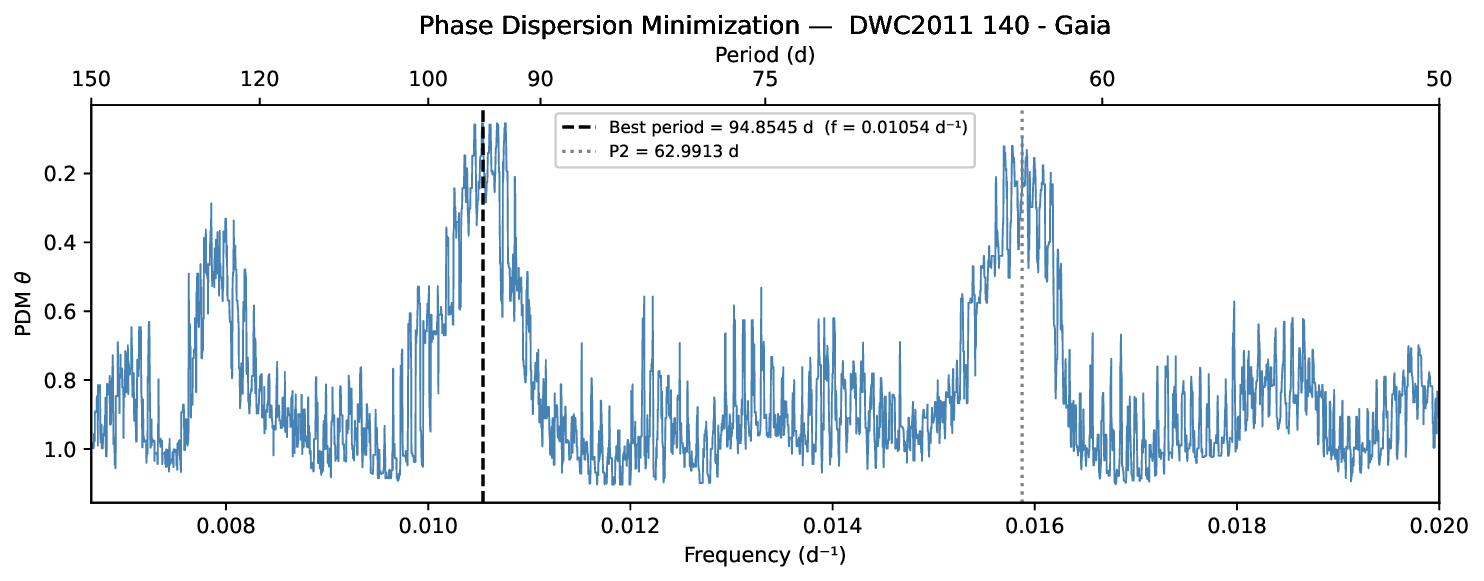}}
\resizebox{1\hsize}{!}{\includegraphics[angle=0]{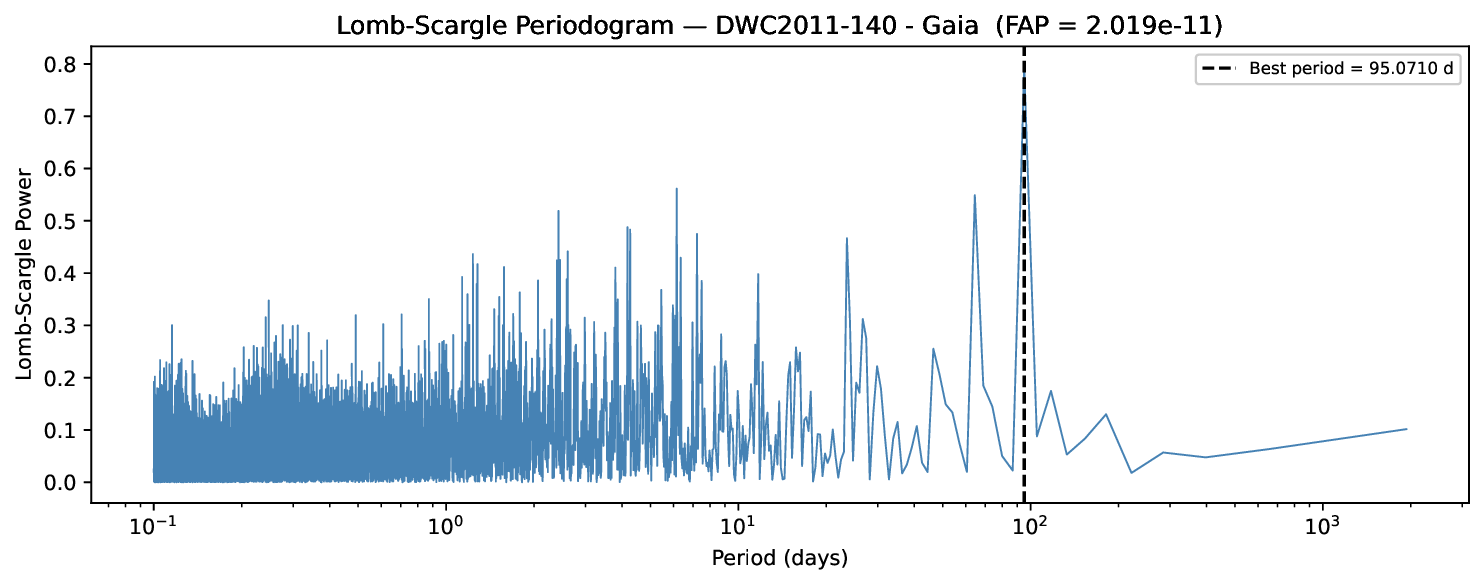}}
\resizebox{1\hsize}{!}{\includegraphics[angle=0]{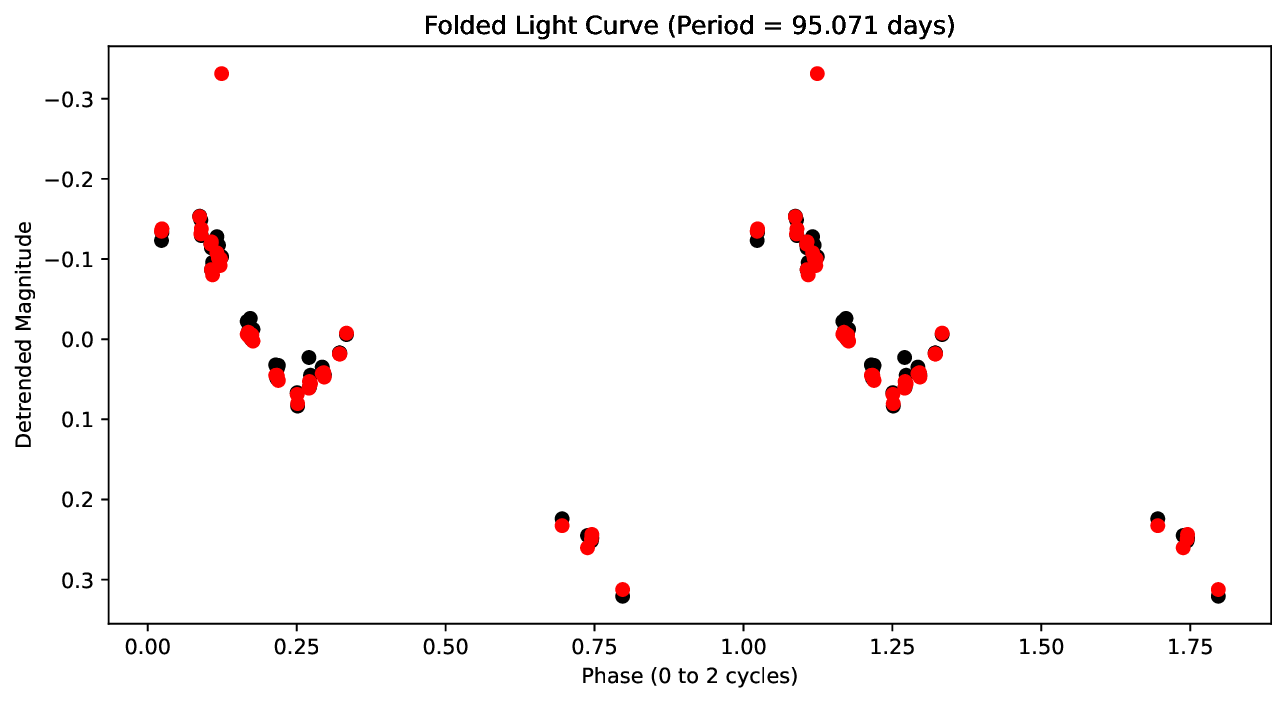}}
\resizebox{1\hsize}{!}{\includegraphics[angle=0]{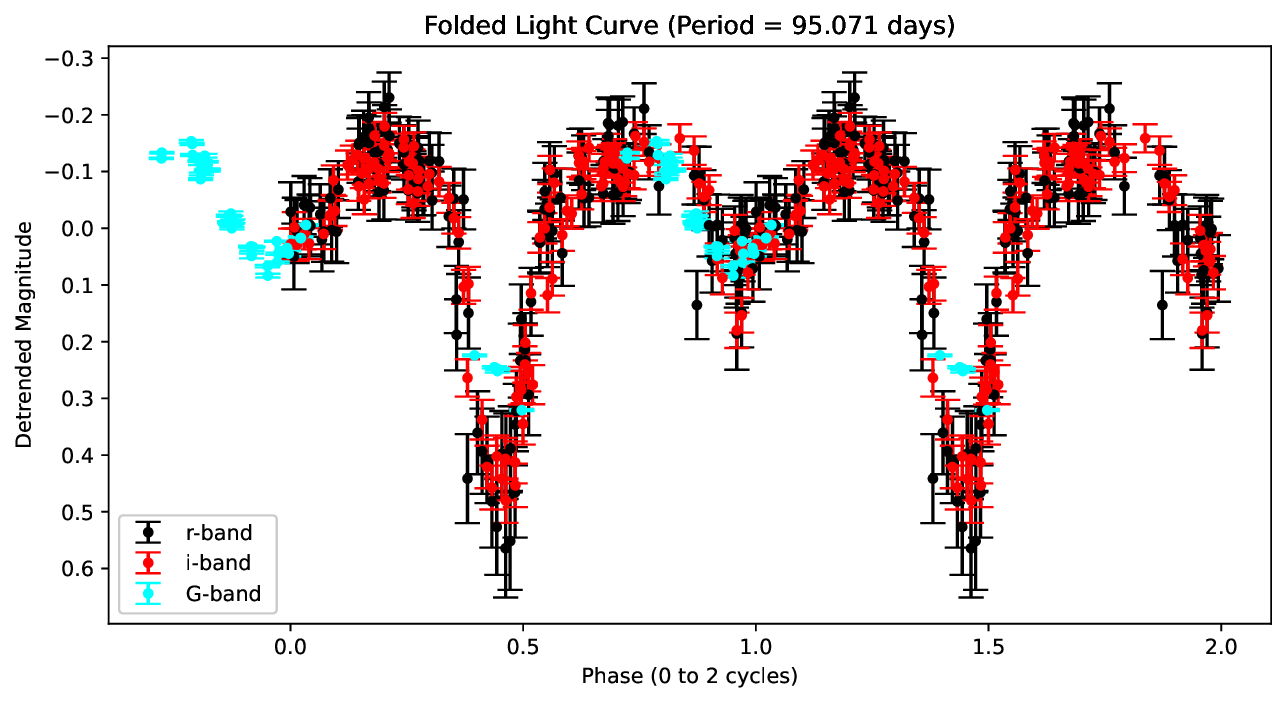}}
\end{centering}
\caption{\label{DWC2011-140} DWC2011\,140. {\it First panel:}
\textit{Gaia} DR3 $G$-band PDM periodogram of  DWC2011\,140. The dashed black line marks the 
highest-probability frequency.
{\it Second panel:}
\textit{Gaia} DR3 $G$-band LS periodogram of  DWC2011\,140. 
The dashed black line marks the 
highest-probability frequency.
{\it Third panel:} \textit{Gaia} DR3 phase diagram of DWC2011\,140.
Filled black circles mark the $G$-band data points, while filled red circles
mark the $RP$ datapoints. The best period adopted is marked on the figure title.
{\it Fourth panel:}  \textit{Gaia} DR3 $G$-band phase diagram (cyan)
is superimposed on the Bochum survey phase diagram 
($r$ band is  black and $i$ band is red). The Bochum phases are calculated
with the \textit{Gaia} DR3 period. The phase of the \textit{Gaia} data is 
shifted by $-0.3$.
}
\end{figure}

\begin{figure}[!]
\begin{centering}
\resizebox{1\hsize}{!}{\includegraphics[angle=0]{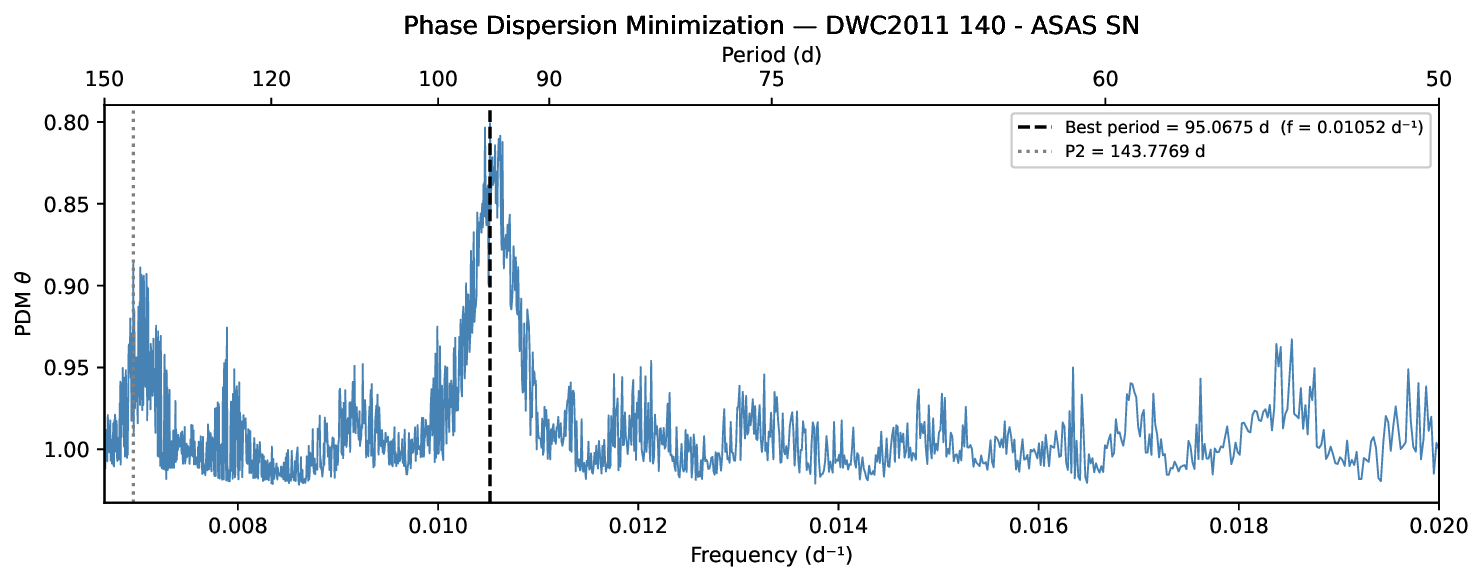}}
\resizebox{1\hsize}{!}{\includegraphics[angle=0]{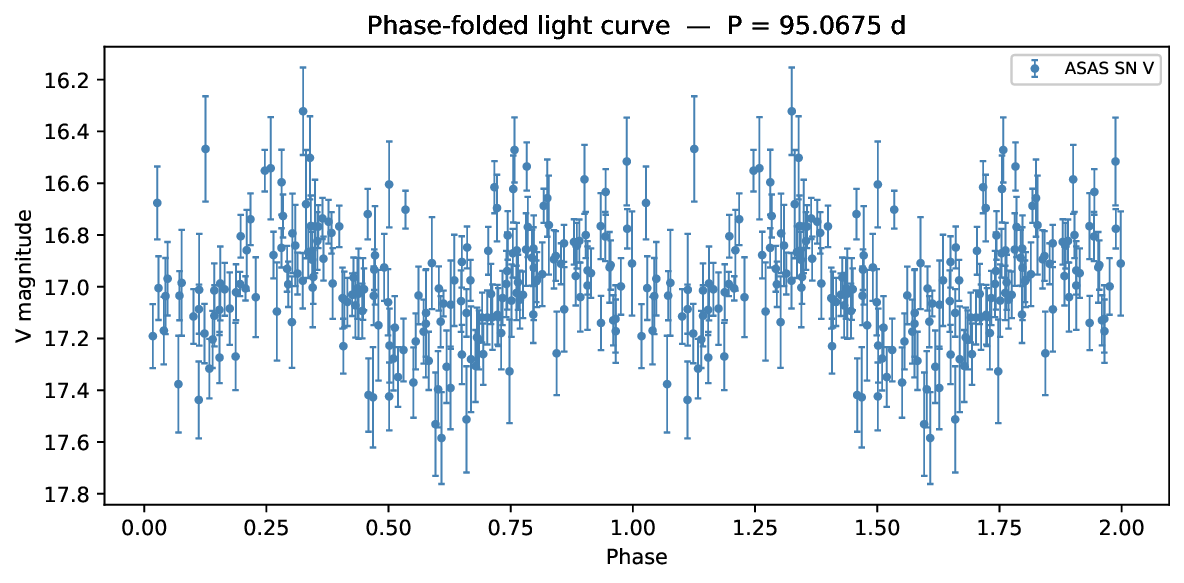}}
\end{centering}
\caption{\label{DWC140-per2} DWC2011\,140. {\it Upper panel:}
$V$-band PDM periodogram of the ASAS\,SN time series of DWC2011\,140. 
The dashed black line marks the first highest-probability frequency.
{\it Lower panel:} ASAS\,SN $V$-band phase diagram of DWC2011\,140.
}
\end{figure}

The \textit{Gaia} DR3 catalog lists only P=0.015 d for DWC2011\,140;
the ASAS\,SN catalog gives P=47.74 d 
\citep[VIZIER II/366/catv2021,][]{jayasinghe19b}.
From the LS periodogram of the \textit{Gaia} and the ASAS\,SN 
light curve of DWC2011\,140, P=95.07 d  with FAP=2.02e-11 
and P= 46.42 d with FAP = 9.174e-05.
The PDM periodograms yield consistent periods, 
P=94.85 d (\textit{Gaia}) and 95.07 d (ASAS\,SN),
as shown in Figs. \ref{DWC2011-140} and \ref{DWC140-per2}.
We adopt P=95.07 d.

For GKF2010\,MN51 no period is listed in \textit{Gaia} DR3, and
the star is not listed in the catalog of ASAS\,SN 
\citep[VIZIER II/366/catv2021,][]{jayasinghe19b}. 
An ASAS\,SN light curve could be retrieved, but the period
derived disagrees with that derived from \textit{Gaia} data.

\section{ Discussion: Periodic variables among the cLBVs}
\label{sect.periodic}

The two robust periodic cLBVs, 
OMN2000\,LS1 and DWC2011\,140, 
should be interpreted cautiously. Periodic light curves with 
primary and secondary minima are not diagnostic of classical 
LBV S\,Doradus behavior. They are more naturally explained by 
binary modulation, eclipses, ellipsoidal variability, or a 
combination of binary light and circumstellar emission. 
Thus, the detection of periodicity does not confirm these 
objects as LBVs. Instead, it identifies them as high-priority 
targets for radial-velocity monitoring and
higher-resolution spectroscopy.

The Bochum light curve of DWC2011\,140 displays clear signatures 
of an eclipsing binary (Fig.\,\ref{DWC2011-140}), 
with  P$ \approx$95\,d 
and well-defined primary and secondary minima ($\Delta\approx0.4$\,mag). 
A secondary minimum is also visible (though noisier) in the 
ASAS\,SN $V$-band light curve ($\Delta\approx0.2$\,mag; 
Fig.\,\ref{DWC140-per2}).
The $K$-band spectrum of DWC2011\,140 shows 
CO emission \citep{geballe19}, a feature more commonly associated 
with the circumstellar disks of B[e]SGs \citep{kraus19}. 
Nevertheless, the star was classified as a cLBV on the basis of its 
spectral variability \citep{clark21}. As shown  
in Fig.\,\ref{fig.lum-teff}, DWC2011\,140 is faint and 
lies well below the lower (hot) edge of the LBV minimum strip. 

The ZTF light curve of OMN2000\,LS1 also reveals a clear 
secondary minimum with maxima asymmetry, which is also consistent 
with a binary system. 
This star is also not extraordinarily luminous; 
it sits precisely on the edge of the LBV minimum strip 
(Fig.\,\ref{fig.lum-teff}).

Recent literature on M31, M33 galaxies, and Magellanic Clouds 
has identified two distinct populations among 
LBVs:  the classical LBVs, 
a high-luminosity group lying above the Humphreys–Davidson limit, 
and a less-luminous group \citep{humphreys16}. The latter can evolve through a 
red-supergiant phase and return blueward, while the classical 
LBVs do not (and historically have been considered as a single star evolution). 
The faintest members of this lower-luminosity population often exhibit
properties overlapping with B[e] supergiants (B[e]SGs), 
for example, they are more isolated.
In M33, massive LBV and/or cLBVs are more clustered and 
distributed as other stars more 
massive than 20 \Msun\
\citep{humphreys16,kostenkov25}; 
while the distribution of iron-stars, emission-line stars
with  \ion{Fe}{ii} in emission -- including LBVs and B[e] stars -- is similar to that 
of  OB stars less massive than 20 \Msun\ (more dispersed), 
and  high-mass X-ray binary (HMXB) donors \citep{kostenkov25}.
However, in the Milky Way, when we plot the 
luminosity–temperature distribution of confirmed binaries versus 
apparently single LBVs and/or cLBVs  (Fig.\,\ref{fig.lum-teff}), 
there is no evidence of separation analogous to 
that reported in M33; the two populations 
(binary and non-binary) overlap significantly, 
preventing a definitive statement on the basis of position in the HR diagram alone. 
This result reinforces the broader conclusion that \textit{Gaia} DR3 is 
effective in flagging unusual variables among cLBVs, 
but it cannot distinguish LBV evolution 
from binary contamination or B[e]SG-like systems. 
The faint cLBVs with periodic variability, DWC2011\,140 and OMN2000\,LS1, 
may belong to the lower-luminosity LBV group.

\citet{mahy22} found that $\approx$70\% are in multiple systems. 
This growing body of evidence has substantially 
revised the classical single-star picture of LBV formation and evolution. 
The three confirmed LBVs  $\eta$\,Carinae\footnote{ $\eta$\,Carinae  is reported 
as a binary in both \citet{lobel25} and \citet{mahy22}; 
more precisely, it is a long-period binary candidate whose 
companion has never been directly detected \citep{weigelt07, damineli96}.}, 
HR\,Carinae, and WRAY\,15-751, 
together with the two cLBVs MWC\,314 and HD\,326823, are all binaries 
or binary candidates \citep[e.g.,][]{lobel25, mahy22}. 
Interferometric observations have additionally revealed 
companions around several other LBVs and cLBVs (visual binaries) 
\citep[$\eta$\,Carinae, P\,Cygni,  HR\,Carinae,  AG\,Carinae, 
W\,243, HD\,160529, 
MN\,46, HD\,326823,][]{mahy22}.

Most classical LBVs  are thought to be 
either wide binaries or appear single, 
with companions too distant to produce eclipses.
Well-documented examples of eclipses include the triple system HD\,5980 in the 
SMC (B1.5\,Ia + WN4--6 + O;  \citealt{moffat98,koenigsberger14}) and the Galactic 
Ofpe/WN9 eclipsing binary GCIRS\,16SW ($P=97$\,d; \citealt{martins06}). 
Partial eclipses have  been reported for HD\,326823 \citep{richardson11} 
and the cLBV MWC\,314 \citep{richardson16, lobel13}. The LMC star R81, 
once thought to be an LBV in an eclipsing binary \citep{tubbesing00}, 
was recently reclassified as a pure binary after its light curve showed 
purely orbital (not wind-driven) signatures \citep{martin25}.
A perfectly regular period and constant amplitude  are instead 
characteristic of binary motion and effectively 
rule out an intrinsic LBV variable
-- the phase-diagrams of DWC2011\,140 and OMN2000\,LS1
appear regular, however, line variations and P\,Cygni profiles
are detected in their spectroscopic observations.

Despite the rapid increase in detected eclipsing binaries 
from TESS ($\gtrsim$5000 new systems), eclipsing supergiants 
remain rare \citep{prisa22}. Among the far more numerous Be stars, 
only V658\,Carinae is a confirmed eclipsing Be+OB binary from TESS data 
\citep{deAmorim25}. \citet{kraus19} list two B[e]SGs in eclipsing 
binaries in M31 and one (GG\,Carinae) in the Galaxy; 
the variability of GG\,Carinae arises also from disk–disk and binary and/or wind 
interactions rather than pure geometric eclipses \citep{porter21, porter22},
and presents asynchronous variations.

\begin{figure}[!]
\begin{center}
\resizebox{.99\hsize}{!}{\includegraphics[angle=0]{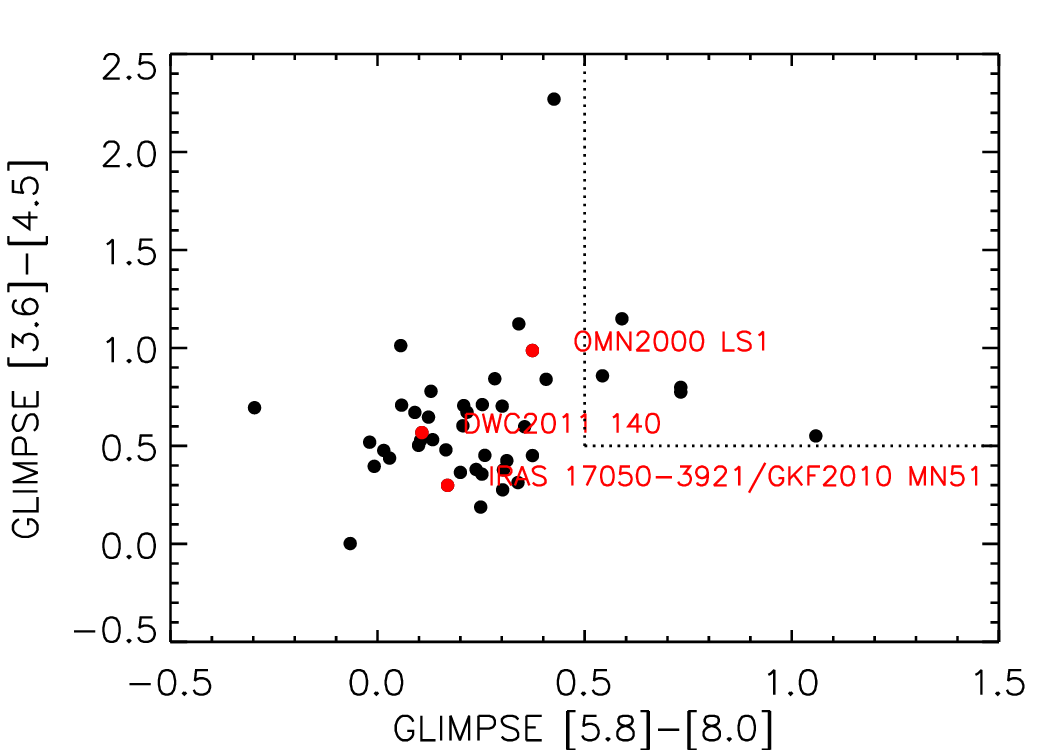}}
\end{center}
\caption{\label{fig.color} 
GLIMPSE $[5.8] - [8.0]$ vs. GLIMPSE $[3.6] - [4.5]$
plot of LBVs and cLBVs. B[e] stars usually fall in the dotted box 
($[5.8] - [8.0]> 0.5 $ mag  $[3.6] - [4.5]>0.5$ mag \citep{kalari26}.
Filled red circles and labels
(OMN2000\,LS1, DWC2011\,140, and GKF2010\,MN51) 
mark the  periodic variables reported  in this work, with 
OMN2000\,LS1, DWC2011\,140 being eclipsing binaries.
}
\end{figure}

Dusty LBVs are mostly associated with cold dust in their outer shells
($\approx 100$\,K) that produces a strong excess at wavelengths 
longer than 10\,$\mu$m \citep{pasquali02}, 
whereas many B[e] supergiants exhibit warm dust (300--1000\,K) that creates 
a prominent mid-infrared bump longward of 1\,$\mu$m, 
between 3 and 8\,$\mu$m \citep{kalari26,messineo20a,clark99,humphreys17}.
Indeed, LBVs and B[e] stars  are characterized 
by different GLIMPSE colors \citep{kalari26}, because the disky B[e] 
are redder.  The GLIMPSE colors of known LBVs
are bluer than  Kalari's area populated by B[e]SG ($[3.6]-[4.5]>0.5$ mag
and $[5.8]-[8.0] >0.5$) and
the three detected periodic variables also
do not appear as red  as B[e]SG stars,  
as shown in Fig.\ \ref{fig.color} --
(0.3 mag ,0.17 mag), (0.57 mag , 0.11 mag), (0.99, 0.37 mag). 
WISE $W3$ and/or $W4$ measurements of the eclipsing
binary are missing\footnote{
A small report on WISE colors of LBVs 
and B[e] stars with infrared extinction-free colors 
to identify selection criteria is in preparation \citep[see also][]{messineo12}.
}.

In summary, DWC2011\,140 and OMN2000\,LS1 are two candidate 
Galactic transitional objects with emission lines characterized 
by P\,Cygni profiles; their light curves display clear primary and 
secondary minima. High-resolution spectroscopic monitoring 
to obtain radial-velocity time series is essential to 
(i) confirm the presence of a companion, 
(ii) derive orbital parameters, 
and (iii) disentangle any variable atmospheric features intrinsic to the primary.
Eclipsing binaries remain keystones for constraining evolutionary and 
dynamical models because they yield direct stellar masses and radii.

\section{Summary}
\label{summary.sect}

We have assembled a vetted sample of 75 Galactic LBVs and cLBVs 
and combined infrared photometry, \textit{Gaia} DR3 source information, 
\textit{Gaia } XP spectra, and time-domain photometry from \textit{Gaia} DR3, 
Bochum, ASAS\,SN, and ZTF. The main conclusions are as follows.

\begin{itemize}
\item[ 1.] \textit{Gaia} DR3 light curves are useful for identifying variability 
among LBVs and cLBVs, but the 2.8-yr baseline is too short to confirm 
most LBVs through S\,Doradus cycles and far too short to sample giant 
eruptions. \textit{Gaia} variability alone is therefore insufficient for 
LBV confirmation.
\item[ 2.] Confirmed Galactic LBVs show a wide range of \textit{Gaia} amplitudes. 
Only a subset exceed about 0.6 mag in $RP$ during the \textit{Gaia} DR3 interval. 
This threshold is useful for selecting cLBVs for follow-up, 
but it is not a classification boundary.
\item[ 3.] Five cLBVs show Gaia $RP$ variations larger than about 0.6 mag. 
Two additional or overlapping cases are highlighted by the Bochum survey. 
These objects deserve further spectroscopic and photometric monitoring, 
especially where the light curves are nonperiodic and resemble slow 
LBV-like changes.
\item[ 4.] OMN2000\,LS1 and DWC2011\,140 show robust periodicity, 
with P=80.29 d and 95.07 d, respectively. Their folded light 
curves show features consistent with binary modulation, including primary 
and secondary minima. These stars should therefore be treated as binary 
candidates rather than as newly confirmed classical LBVs.
\item[ 5.] The infrared catalog remains essential supporting material 
because it validates source identifications in crowded and reddened 
Galactic-plane fields and provides circumstellar-dust context for 
separating cLBVs from B[e]SG-like or otherwise contaminated objects.
The catalog is made public.
\item[ 6.] Future progress requires long-baseline monitoring that 
\textit{combines photometry, color changes, spectroscopy, and radial velocities. 
Gaia}, ASAS\,SN, ZTF, and future LSST observations will be powerful discovery 
tools, but confirmation of LBV status will remain a historical and 
multidiagnostic problem.

\end{itemize}

\section{Data availability}
An atlas of the XP spectra of LBV and cLBVs is available at
\href{https://mariamessineo.github.io/lbv_bprp_dr3/index.html}{lbv\_bprp\_dr3\_github}.
An atlas of infrared finding charts is also available at the same link,
\href{https://mariamessineo.github.io/lbv_bprp_dr3/index.html}{lbv\_bprp\_dr3\_github}.

\begin{acknowledgements}
This work has made use of data from the European Space Agency (ESA) mission {\it Gaia}
($http://www.cosmos.esa.int/gaia$), processed by the {\it Gaia} Data Processing and Analysis
Consortium (DPAC, $http://www.cosmos.esa.int/web/gaia/dpac/consortium$). Funding for the DPAC
has been provided by national institutions, in particular the institutions participating in the {\it
Gaia} Multilateral Agreement. 
This publication makes use of data products from the Two Micron All
Sky Survey, which is a joint project of the University of Massachusetts and the Infrared Processing
and Analysis Center / California Institute of Technology, funded by the National Aeronautics and Space
Administration and the National Science Foundation. 
This work is based on observations made with
the Spitzer Space Telescope, which is operated by the Jet Propulsion Laboratory, California
Institute of Technology under a contract with NASA.
This research made use of data products from the Midcourse Space Experiment, the processing of which
was funded by the Ballistic Missile Defense Organization with additional support from the NASA
office of Space Science. 
This publication makes use of data products from WISE, which is a joint
project of the University of California, Los Angeles, and the Jet Propulsion Laboratory / California
Institute of Technology, funded by the National Aeronautics and Space Administration. 
We thank Martin Haas for publishing the Bochum survey.
ASAS\,SN is supported by the Gordon and Betty Moore Foundation through 
grant GBMF5490 to the Ohio State University, NSF grants AST-1515927 
and AST-1908570, the Alfred P. Sloan Foundation grant G-2021-14192, 
the Mt. Cuba Astronomical Foundation, the Center for Cosmology and 
AstroParticle Physics at the Ohio State University, the Chinese 
Academy of Sciences South America Center for Astronomy (CASSACA), 
the Villum Foundation, and George Skestos. 
ASAS\,SN is also hosted in part by Las Cumbres Observatory.
Based on observations obtained with the Samuel Oschin Telescope 
48-inch and the 60-inch Telescope at the Palomar Observatory 
as part of the Zwicky Transient Facility project. ZTF is supported 
by the National Science Foundation under Grants No. AST-1440341 
and AST-2034437 and a collaboration including current partners 
Caltech, IPAC, the Oskar Klein Center at Stockholm University, 
the University of Maryland, University of California Berkeley, 
the University of Wisconsin at Milwaukee, the University of 
Warwick, Ruhr University, Cornell University, Northwestern 
University, and Drexel University. Operations are conducted by COO, IPAC, and UW.
This research has made use of the VizieR catalog access tool, 
CDS, Strasbourg, France, and SIMBAD database.
This research utilized  the NASA’s Astrophysics Data System Bibliographic Services.\\ 
We thank the Python, Astropy, NumPy, Pandas, Pivo, SciPy, Matplolib 
communities for providing their packages.  
This catalog is dedicated to the memory of our esteemed colleagues
Dr. Simon Clark, Dr. Vasilii Gvaramadze, and Dr. Karl Menten.
The authors are infinitely grateful to Dr. Paco Najarro
for providing input on the list of collected LBVs.
We thank the anonymous referees and the editor of A\&A for their advice and patience.
\end{acknowledgements}

\clearpage
\begin{appendix}
\section{Luminosity of known LBV}
\label{appendixlum}

Luminosities and temperatures of LBVs were collected from the literature, 
and listed in Table\,\ref{table.lumin}. When Gaia DR3 parallaxes are available 
(see Sect. \ref{infrared}),  Gaia distances by \citet{bailer21} are 
retained and the luminosities are rescaled.
 We are aware of the new publication by \citet{bernini26},
that we saw while submitting.
Our compilation contains some stars that are not in the new work of 
by \citet{bernini26}.

\begin{table*}
\caption{\label{table.lumin} Luminosities and temperatures 
of LBVs and cLBVs. The values are collected from the literature. }
\begin{tabular}{llrrrrrrrl}
\hline
\hline
                    Name &   Classi- &  Lum1 &  Temp1 &  Lum2 &  Temp2 &     DM(lit) & DM & $\varpi / \sigma_\varpi$ &                                         Reference \\
                         &   fication &       & [K]    &       & [K]    & [mag] &[mag] \\
\hline
                   $\eta$\,Carinae  &     LBV &   6.70 &  35200 &   6.70 &  $..$ &  11.81 &      $..$ &      $..$ &                                    $16$,  $18$ \\
                    P\,Cygni &     LBV &   5.87 &  20500 &   5.67 &  18500 &  11.02 & 10.95$^{0.22}_{-0.21}$ &     7.73 &                                    $8$,  $1$ \\
                   HR\,Carinae  &     LBV &   5.60 &  19000 &   5.50 &   9000 &  13.31 &      $..$ &      $..$ &                                    $8$,  $1$ \\
                   AG\,Carinae  &     LBV &   6.17 &  24000 &   5.77 &   7500 &  13.55 & 13.48$^{0.20}_{-0.18}$ &     9.22 &                       $41$, $34$, $8$,  $1$ \\
WRA15-751/IRAS\,11065$-$6026 &     LBV &   6.10 &  28840 &   5.60 &   8413 &  13.89 & 14.40$^{0.25}_{-0.26}$ &     4.28 &                                        $26$  \\
           GKM2012\,WS1 &     LBV &   4.50 &  22000 &   4.50 &  12000 &  14.39 & 14.39$^{0.25}_{-0.21}$ &     2.79 &                       $10$,  $3$, $8$ \\
              WRAY\,16-137 &     LBV &   6.05 &  18000 &   6.05 &  $..$ &  12.65 & 12.66$^{0.35}_{-0.27}$ &     4.55 &                                   $14$, $3$ \\
           GKF2010\,MN44 &     LBV &   6.40 &  12000 &   6.40 &  $..$ &  15.88 &      $..$ &      $..$ &                          $20$,  $11$,  $3$ \\
                    W243 &     LBV &   6.30 &  18500 &   6.10 &   8000 &  14.37 & 14.39$^{0.60}_{-0.81}$ &     0.10 &                                    $8$,  $1$ \\
           GKM2012\,MN48 &     LBV &   5.50 &  15000 &   5.40 &  13000 &  11.53 & 13.52$^{0.66}_{-0.59}$ &     1.33 &                                    $8$,  $1$ \\
               WRAY\,17-96 &     LBV &   6.25 &  13000 &   6.25 &  $..$ &  13.27 & 13.29$^{0.47}_{-0.57}$ &     1.51 &                                            $33$ \\
                HD\,160529 &     LBV &   5.29 &  11500 &   5.19 &   7000 &  11.25 & 11.23$^{0.16}_{-0.12}$ &    14.55 &                                    $8$,  $1$ \\
            G0.120$-$0.048 &     LBV &   6.60 &  12000 &   6.60 &  $..$ &  14.52 &      $..$ &      $..$ &                     $22$,  $13$,   $10$ \\
                  FMM362 &     LBV &   6.20 &  11300 &   6.20 &  $..$ &  14.52 &      $..$ &      $..$ &        $36$, $22$,  $24$,  $10$ \\
                  MWC\,930 &     LBV &   5.60 &  22500 &   5.60 &  $..$ &  12.39 & 14.27$^{0.71}_{-0.56}$ &     1.84 &                                  $29$ \\
             G24.73+0.69 &     LBV &   5.60 &  12000 &   5.60 &  $..$ &  13.58 & 13.66$^{1.18}_{-0.61}$ &     0.61 &                                            $32$ \\
                AFGL2298 &     LBV &   6.20 &  11700 &   6.30 &  10300 &  15.00 &      $..$ &      $..$ &                                  $31$,  $23$ \\
              SBW20071 &    cLBV &   5.29 &  21000 &   5.29 &  $..$ &  14.23 & 13.69$^{0.14}_{-0.10}$ &     8.41 &                                            $8$ \\
                HEN3-519 &    cLBV &   5.58 &  27500 &   5.58 &  $..$ &  14.52 & 14.32$^{0.28}_{-0.26}$ &     5.75 &                               $43$,  $42$ \\
              WRAY\,15-906 &    cLBV &   5.63 &  25000 &   5.63 &  $..$ &  12.74 & 13.31$^{0.21}_{-0.22}$ &     7.36 &                            $6$,  $9$  \\
          IRAS\,16115-5044 &    cLBV &   5.75 &  11000 &   5.75 &  $..$ &  12.83 & 12.42$^{0.51}_{-0.67}$ &     1.62 &                    $7$,  $5$, $9$ \\
               HD\,326823  &    cLBV &   5.10 &  22000 &   4.90 &  20000 &  10.70 & 10.71$^{0.05}_{-0.05}$ &    29.32 &                       $27$,  $8$,  $1$ \\
              Pistol Star &    cLBV &   6.20 &  11800 &   6.20 &  $..$ &  14.52 &      $..$ &      $..$ &                                         $24$ \\
                HD\,316285 &    cLBV &   5.83 &  15000 &   5.83 &  $..$ &  11.34 & 12.32$^{0.17}_{-0.18}$ &     8.86 &                                         $38$ \\
             LBV 1806-20 &    cLBV &   5.70 &  24000 &   5.70 &  $..$ &  14.70 &      $..$ &      $..$ &                                 $30$,   $25$ \\
             LBV 1813-178 &    cLBV &   5.32 &  16000 &   5.32 &  $..$ &  13.41 & 12.01$^{0.32}_{-0.29}$ &     3.04 &                                        $17$ \\
               G23\_CCLBV &    cLBV &   5.06 &  13200 &   5.06 &  $..$ &  13.31 & 12.25$^{0.70}_{-0.48}$ &     1.64 &                                        $15$ \\
               G25.5+0.2 &    cLBV &   5.90 &  34000 &   5.90 &  $..$ &  15.81 &      $..$ &      $..$ &                                    $35$,  $32$ \\
             G26.47+0.02 &    cLBV &   6.00 &  17000 &   6.00 &  $..$ &  14.06 &      $..$ &      $..$ &                                            $32$ \\
                  MWC\,314 &    cLBV &   6.31 &  25000 &   6.31 &  $..$ &  12.39 & 12.92$^{0.18}_{-0.17}$ &     9.06 &                                  $39$ \\
                 OMN2000\,LS1 &    cLBV &   5.30 &  13400 &   5.30 &  $..$ &  12.66 & 13.27$^{0.62}_{-0.42}$ &     1.34 &                                           $23$  \\
          GKF2010\,MN112 &    cLBV &   6.18 &  17400 &   6.18 &  $..$ &  15.66 &      $..$ &      $..$ &                                 $19$, $2$ \\
             G79.29+0.46 &    cLBV &   5.40 &  20400 &   5.40 &  $..$ &  11.15 & 11.15$^{0.23}_{-0.23}$ &     5.35 &  $40$,   $37$,  $21$,  $12$ \\
          DWC2011\,140 & cLBV &   4.89 &  21500 &   4.89 & $..$  &   $..$ & 10.92$^{0.25}_{-0.20}$   & 5.49 & \\
\hline
\end{tabular}
\begin{list}{}
\item -- Lum1, Temp1, Lum2, Temp2, DM(Lit) are collected from the listed references. When 
the Gaia DR3 $\varpi / \sigma_\varpi$ is larger than 4,
Lum1 and Lum2 were corrected with the  Gaia DR3 distances by \citet{bailer21}.
\item -- DMlit= DM values collected from the literature.
\item -- DM= Distance moduli derived with the distances by \citet{bailer21},
only positive parallaxes are considered.
\item -- $\varpi / \sigma_\varpi$ this is the inverse of the fractional error of the parallax.

\item -- Reference: 
$[1]$=\citet{mahy22};
$[2]$=\citet{maryeva22};
$[3]$=\citet{bailer21};
$[4]$=\citet{clark21};
$[5]$=\citet{figer20};
$[6]$=\citet{maryeva20};
$[7]$=\citet{figer20};
$[8]$=\citet{smith19};       
$[9]$=\citet{kniazev15b};
$[10]$=\citet{kniazev15};
$[11]$=\citet{gvaramadze15};  
$[12]$=\citet{agliozzo14};    
$[13]$=\citet{lau14};  
$[14]$=\citet{gvaramadze14};  
$[15]$=\citet{messineo14};
$[16]$=\citet{naze12};        
$[17]$=\citet{messineo11};
$[18]$=\citet{Groh10};
$[19]$=\citet{gvaramadze10b}; 
$[20]$=\citet{gvaramadze10};  
$[21]$=\citet{jimenez10};     
$[22]$=\citet{mauerhan10};    
$[23]$=\citet{clark09a};       
$[24]$=\citet{najarro09};
$[25]$=\citet{bibby08};
$[26]$=\citet{sterken08};
$[27]$=\citet{marcolino07};
$[28]$=\citet{clark05};
$[29]$=\citet{miroshnichenko05};
$[30]$=\citet{figer04};
$[31]$=\citet{clark03b};  
$[32]$=\citet{clark03};  
$[33]$=\citet{egan02};
$[34]$=\citet{Stahl01};  
$[35]$=\citet{clark00}; 
$[36]$=\citet{geballe00};
$[37]$=\citet{voors00};
$[38]$=\citet{hillier98};
$[39]$=\citet{miroshnichenko98};
$[40]$=\citet{trams98};
$[41]$=\citet{humphreys94}; 
$[42]$=\citet{smith94b};
$[43]$=\citet{davidson93}.
\item -- WRAY\,16-137 is similar to P\,Cygni, and a \Teff=18000 K is assumed. 
It has E(B-V)=3.7 mag, \Av=11.5 mag and DM+Mv=3.8 mag  \citet{gvaramadze14},
By considering the Gaia distance by \citet{bailer21} 
and a BC\_V=$-1.54$, like that of P\,Cygni, the luminosity was estimated.

\item -- For GKF2010\,MN44, several possible distances and luminosity
are tabulated by \citep{gvaramadze15}.  DM = 15.88 mag is taken,
because it is closer to the DM=15.59 mag 
by \citet[of poor quality though][]{bailer21}.

\item --  For DWC2011\,140, we estimated a bolometric magnitude 
Mbol=$-7.49$ mag by assuming
a B0-2 I spectral type \citep{mauerhan10}, 
intrinsic colors $J-Ks$=$-0.09$ mag and $H-Ks$=$-0.03$ mag, 
and a bolometric correction BC$_K$=$-2.65$ mag \citep{messineo11},
2MASS $J= 8.71$ and $Ks=7.03$ mag,  
the distance moduli DM=10.92 mag \citep{bailer21}, and by estimating
the interstellar extinction as \Aks=(2MASS $J$-\Ks+0.09)*0.537 mag = 0.95 mag
\citep{messineo05}.

\end{list}
\end{table*}

\section{Table of Gaia DR3  counterparts and photometric variations}
\label{amplitudes.table}

The 14 Galactic LBVs and 36 cLBVs with a Gaia DR3  counterpart
are listed in Table\,\ref{lbvlist}.
The photometric variations are estimated as the maximum minus 
minimum magnitudes from the Gaia DR3, ASAS\,SN, and Bochum light curves.

\begin{sidewaystable*}\renewcommand{\arraystretch}{0.9}
\vspace*{16.5cm} 
\caption{ \label{lbvlist} List of the 14  Galactic LBVs and 36  cLBVs with Gaia DR3 
counterparts.}
\resizebox{1\textheight}{!}{   
\begin{tabular}{@{\extracolsep{-.05in}}ll|llllll|ll|cl|llllll}
\hline
\hline
                     NAME &         twomass  	 &           \multicolumn{6}{c}{Gaia DR3}         & \multicolumn{2}{c}{Bochum}&  \multicolumn{2}{c}{ASAS\,SN} & \multicolumn{5}{c}{Comments}  \\
                          &                      &   source\_id     & RUWE & snrBIN&$\Delta$ G & $\Delta$ BP & $\Delta$ RP & $\Delta$R & $\Delta$I & $\Delta$V$^c$ &  type                    & Bin&Xray  &  BP/RP       & IR\_neb & neb\_lob     \\
                          &                      &                  &      &       &[mag]      &  [mag]      &[mag]        & [mag]     &[mag]      &[mag]      &                          &     &  &   \ion{H\kern-0.2em}{$\alpha$}  &          & \\
\hline
                               $\eta$\,Carinae  & 10450360$-$5941040 & 5350358584482202880 &   $..$ &           $..$ &   0.20 &    0.18 &    0.13 &     .. &     .. &       $..$ &    .. & sp/int &    X &     .. &  .. & large bipolar \\
                                P\,Cygni & 20174719+3801585 & 2061242908036996352 &  0.92 &          1.41 &   0.12 &    0.12 &    0.19 &     .. &     .. &    (0.75) &  SDOR &    int &   X2 &     X* &  .. &     spherical \\
\hdashline
                                HR\,Carinae  & 10225385$-$5937283 & 5255045082580350080 &  0.94 &         17.39 &   0.22 &    0.19 &    0.26 &   0.25 &   0.00 &    (0.42) &  SDOR & sp/int &   .. &     X* &  .. &       bipolar \\
                                AG\,Carinae  & 10561158$-$6027128 & 5338220285385672064 &  1.04 &          1.17 &   1.18 &    1.08 &    1.29 &   0.65 &   0.00 &    (1.70) &  SDOR &    int &   X2 &      X &   X & non$-$spherical \\
          WRAY\,15-751/IRAS\,11065$-$6026 & 11084006$-$6042516 & 5337309477433273728 &  0.92 &           $..$ &   0.59 &    0.50 &    0.59 &     .. &     .. &      0.80 &  SDOR &     sp &   .. &      $-$ &  .. &       bipolar \\
                          GKM2012\,WS1 & 13362862$-$6345387 & 5864989022016713728 &  1.16 &           $..$ &   1.15 &    1.02 &    1.15 &   1.46 &   1.56 &      1.25 & SDOR: &     .. &   .. &     X* &  .. &               \\
                           WRAY\,16$-$137 & 13501536$-$6148552 & 5865577604279098112 &  0.82 &           $..$ &   0.81 &    0.61 &    0.86 &   1.61 &   1.73 &      0.41 &  SDOR &     .. &   .. &      X &  .. &               \\
                         GKF2010\,MN44$^b$ & 16323995$-$4942137 & 5940576971716473344 &  1.00 &           $..$ &   0.62 &    0.55 &    0.60 &   1.21 &   1.40 &      0.66 &     L &     .. &    X &      X &   X &               \\
               Cl* Westerlund 1 W 243 & 16470749$-$4552290 & 5940105830990286208 &  1.15 &           $..$ &   0.17 &    0.33 &    0.26 &   0.20 &   0.17 &      0.49 &  SDOR &    int &    X &      $-$ &  .. &               \\
                         GKM2012 MN48 & 16493770$-$4535592 & 5940216130049700480 &  1.16 &           $..$ &   0.31 &    0.29 &    0.30 &     .. &     .. &       $..$ &    .. &     .. &   .. &      $-$ &   X &               \\
                           WRAY\,17$-$96 & 17413543$-$3006389 & 4056941758956836224 &  0.90 &           $..$ &   0.43 &    0.47 &    0.63 &   0.59 &   0.64 &      0.48 &  SDOR & no$-$det &   .. &      X &   X &     spherical \\
                            HD\,160529 & 17415902$-$3330136 & 4053887521876855808 &  0.88 &          3.29 &   0.33 &    0.33 &    0.41 &     .. &     .. &    (3.74) &  SDOR &    int &   X2 &      $-$ &  .. &               \\
                              MWC\,930 & 18262523$-$0713177 & 4159973866869462784 &  0.91 &           $..$ &   0.24 &    0.18 &    0.24 &     .. &     .. &      0.23 &  SDOR &     .. &   X2 &     .. &  .. &     spherical \\
                          G24.73+0.69 & 18335528$-$0658386 & 4255908794692238848 &  1.07 &           $..$ &   0.41 &   $..$ &    0.39 &   0.76 &   0.62 &      1.27 &     L &     .. &   .. &      $-$ &   X &               \\
\hline
                            SWB2007\,1 & 10401937$-$5949099 & 5254478417451126016 &  0.86 &           $..$ &     .. &      .. &      .. &     .. &     .. &       $..$ &    .. &     .. &   .. &      $-$ &  .. &               \\
                            Hen\,3$-$519 & 10535958$-$6026444 & 5338229115839425664 &  1.07 &           $..$ &   0.29 &    0.29 &    0.30 &   0.16 &   0.18 &      0.62 & SDOR: &     .. &   .. &     X* &   X &               \\
                          GKF2010\,MN1 & 11441803$-$6245210 & 5333327905312912768 &  0.95 &           $..$ &     .. &      .. &      .. &     .. &     .. &       $..$ &    .. &     .. &   .. &      X &   X &               \\
                          WRAY\,15$-$906 & 11544353$-$6313311 & 5333080274685374720 &  1.02 &           $..$ &   0.35 &    0.30 &    0.44 &   0.24 &   0.25 &      0.37 &    SR &     .. &   .. &     X* &   X &               \\
                          GKF2010\,MN7 & 13100486$-$6311300 & 5862438288106749056 &  0.95 &           $..$ &     .. &      .. &      .. &     .. &     .. &       $..$ &    .. &     .. &   .. &     X* &   X &               \\
                          GKF2010\,MN8 & 13104384$-$6317457 & 5862421967230819200 &  0.99 &           $..$ &  $..$ &   $..$ &    0.29 &     .. &     .. &       $..$ &    .. &     .. &   .. &      $-$ &   X &               \\
                         GKF2010\,MN30 & 15484207$-$5507422 & 5884424088255200128 &  1.00 &           $..$ &   0.24 &    0.27 &    0.20 &     .. &     .. &       $..$ &    .. &     .. &   .. &      $-$ &   X &               \\
             IRAS\,16115$-$5044 & 16151795$-$5052197 & 5935061172876722176 &  0.98 &           $..$ &  $..$ &   $..$ &    0.97 &     .. &     .. &       $..$ &    .. &     .. &   .. &      $-$ &   X &               \\
                         GKF2010\,MN41 & 16263427$-$5021018 & 5934575497980191872 &  1.29 &           $..$ &   0.30 &   $..$ &    0.31 &     .. &     .. &       $..$ &    .. &     .. &   .. &      $-$ &   X &               \\
       WMD2010\,14/IRAS\,16254$-$4739 & 16290377$-$4746264 & 5941957138026204416 &  1.05 &           $..$ &   0.60 &   $..$ &    0.63 &     .. &     .. &       $..$ &    .. &     .. &   .. &      $-$ &   X &               \\
                GKF2010\,MN46 & 16431636$-$4600424 & 5943135780115239680 &  1.07 &           $..$ &   0.30 &    0.40 &    0.44 &   0.53 &   0.56 &      0.35 &     L &    int &   .. &      X &   X &               \\

\hline
\end{tabular}
}
\begin{list}{\textendash}{}
\item  LBVs are located in the top part of the Table, followed by the cLBVs.
Each group is sorted on the basis of the 2MASS identification number.
\item  The source NAME is followed by the 2MASS ID (twomass),  
the Gaia DR3 source\_id, the Gaia DR3 RUWE parameter, 
the  snrPMaH2EG3b flag (snrBIN) by \citet{kervella22},
the  magnitude variation in the Gaia  $G$-band ($\Delta\,G$), 
the  magnitude variation in the Gaia $BP$-band ($\Delta\,BP$), 
the  magnitude variation in the Gaia $RP$-band ($\Delta\,RP$), 
the  magnitude variation in the Bochum $r'$-band ($\Delta\,R$), 
the  magnitude variation in the Bochum $i'$-band ($\Delta\,I$), 
the  magnitude variation in the ASAS\,SN $V$-band ($\Delta\,V$), 
the variable type given by ASAS\,SN,
a flag reporting detected binary systems by \citet{mahy22}, \citet{lobel25},
and this work (Bin = interferometer (int), eclipsing binary (ecl), 
adaptive optics (AO), and spectroscopy (sp)),
a flag indicating detection at X-ray wavelength (Xray),
a flag indicating detection of  \ion{H\kern-0.2em}{$\alpha$}  emission in the Gaia BP/RP spectrum
(particular strong emission are marked with (*)),
a flag indicating the detection of an extended infrared nebula (IR\_neb), and
a flag reporting the optical nebula listed (neb\_lob) in \citet{lobel25}.
\item  Xray: Xray emission of  $\eta$\,Carinae, W243, and  G26.47+0.02 
are given in the work of \citet{naze12}.
The X2 symbol denotes additional Xray emitters associated with LBVs 
listen by \citet{lobel25} --  AG\,Carinae, HD\,160529,
P\,Cygni, MWC\,930, and MWC\,314; 
all but MWC\,930 are also in VIZIER.\\
In addition, from VIZIER, \citet{fortin23} reports that  Xray emission in
GKF2010\,MN44.
\item (a) Two minima appear in the light curve of HD\,326823 \citep{richardson11};
The light curve of the cLBV MWC\,314 shows also a binary modulation
\citep{richardson16,lobel13}. 
\item (b) GKF2010\,MN44 is listed as an HMXB by \citet{fortin23}.
\item (c)  ASAS\,SN $\Delta$V are recalculated with the same routine
adopted for the Gaia light curves (see text); values are  
between round parenthesis when $V<10$ mag,
because above saturation limit.

\end{list}
\end{sidewaystable*}

\addtocounter{table}{-1} 
\begin{sidewaystable*}\renewcommand{\arraystretch}{0.9}
\vspace*{16.5cm} 
\caption{  Table \ref{lbvlist} continued.  }
\resizebox{1\textheight}{!}{
\begin{tabular}{@{\extracolsep{-.07in}}ll|llllll|ll|cl|llllll}
\hline
\hline
                     NAME &         twomass  	 &           \multicolumn{6}{c}{Gaia DR3}         & \multicolumn{2}{c}{Bochum}&  \multicolumn{2}{c}{ASAS\,SN} & \multicolumn{5}{c}{Comments}  \\
                          &                      &   source\_id     & RUWE & snrBIN&$\Delta$ G & $\Delta$ BP & $\Delta$ RP & $\Delta$R & $\Delta$I & $\Delta$V$^c$ &  type                    & Bin&Xray  &  BP/RP       & IR\_neb & neb\_lob     \\
                          &                      &                  &      &       &[mag]      &  [mag]      &[mag]        & [mag]     &[mag]      &[mag]      &                          &     &  &   \ion{H\kern-0.2em}{$\alpha$}  &          & \\
\hline
                            HD\,326823$^a$ & 17065390$-$4236397 & 5965495757804852992 &  1.08 &           $..$ &     .. &      .. &      .. &   0.20 &   0.21 &    (0.46) & SDOR: & sp/int &   .. &      $-$ &  .. &               \\
GKF2010\,MN51 & 17082913$-$3925076 & 5972983638153011840 &  1.22 &           $..$ &  $..$ &   $..$ &    1.32 &     .. &     .. &       $..$ &    .. &     .. &   .. &     .. &   X &               \\
                         GKF2010\,MN55 & 17110094$-$3945174 & 5972934877369594240 &  1.01 &           $..$ &   0.46 &   $..$ &    0.37 &     .. &     .. &       $..$ &    .. &     .. &   .. &      $-$ &   X &               \\
                         GKF2010\,MN56 & 17111668$-$3931312 & 5972944571113006720 &  1.00 &           $..$ &  $..$ &   $..$ &    0.39 &     .. &     .. &       $..$ &    .. &     .. &   .. &      $-$ &   X &               \\
                          GKM2012 WS2 & 17203175$-$3309490 & 5979217486808094720 &  1.03 &           $..$ &   0.08 &    0.17 &    0.07 &     .. &     .. &      0.23 &     L &     .. &   .. &      X &   X &               \\
                         GKF2010\,MN58 & 17374754$-$3137333 & 4055062727939772800 &  1.22 &           $..$ &     .. &      .. &      .. &     .. &     .. &       $..$ &    .. &     .. &   .. &     .. &   X &               \\
                         GKF2010\,MN64 & 17435981$-$3028384 & 4056684954262144896 &  0.84 &           $..$ &   0.13 &    0.21 &    0.13 &   0.23 &   0.24 &      0.23 &     L & no$-$det &   .. &      $-$ &   X &               \\
                 DWC2011\,140 & 17445945$-$2905258 & 4057105238280140160 &  0.93 &           $..$ &   0.47 &    0.81 &    0.65 &   0.79 &   0.65 &      1.23 &    SR &      ecl &   .. &      X &  .. &               \\
              2MASS\,J17470940$-$2849235 & 17470940$-$2849235 & 4057501990180723712 &  1.10 &           $..$ &     .. &      .. &      .. &     .. &     .. &       $..$ &    .. &     .. &   .. &      X &  .. &               \\
                            HD\,316285 & 17481403$-$2800531 & 4057682692354437888 &  0.87 &          2.59 &   0.13 &    0.16 &    0.16 &     .. &     .. &    (1.81) & SDOR: & no$-$det &   .. &     X* &   X &               \\
                         GKF2010\,MN76 & 18070516$-$2015163 & 4094936822058475136 &  0.98 &           $..$ &     .. &      .. &      .. &     .. &     .. &       $..$ &    .. &     .. &   .. &     .. &   X &               \\
                           MFD2008 15 & 18132098$-$1749469 & 4095731906397582976 &  1.04 &           $..$ &   0.15 &    0.20 &    0.12 &     .. &     .. &       $..$ &    .. &     .. &   .. &      X &  .. &               \\
                         GKF2010\,MN81 & 18334346$-$0823353 & 4156400209244873728 &  1.00 &           $..$ &     .. &      .. &      .. &     .. &     .. &       $..$ &    .. &     .. &   .. &     .. &   X &               \\
            SSTGLMC G023.4521$-$00.0948 & 18342170$-$0828209 & 4156489445780902016 &  1.12 &           $..$ &  $..$ &   $..$ &    0.16 &     .. &     .. &       $..$ &    .. &     .. &   .. &      X &  .. &               \\
                          G26.47+0.02 & 18393224$-$0544204 & 4256472156970283904 &  1.05 &           $..$ &   0.44 &   $..$ &    0.47 &     .. &     .. &       $..$ &    .. &     .. &    X &      X &   X &               \\
                         GKF2010\,MN84 & 18415965$-$0515409 & 4256610287438264192 &  1.03 &           $..$ &   0.18 &   $..$ &    0.15 &     .. &     .. &       $..$ &    .. &     .. &   .. &      $-$ &   X &               \\
                         GKF2010\,MN87 & 18422247$-$0504300 & 4256620045576946688 &   $..$ &           $..$ &     .. &      .. &      .. &     .. &     .. &       $..$ &    .. &     .. &   .. &     .. &   X &               \\
                         GKF2010\,MN96 & 18510295$-$0058242 & 4266044921851779840 &  1.15 &           $..$ &     .. &      .. &      .. &     .. &     .. &       $..$ &    .. &     .. &   .. &     .. &   X &               \\
                        GKF2010\,MN101 & 19062457+0822015 & 4307460432455822976 &  2.48 &           $..$ &   0.58 &   $..$ &    0.45 &     .. &     .. &       $..$ &    .. &     .. &   .. &      X &   X &               \\
                              MWC\,314$^a$ & 19213397+1452570 & 4319930096909297664 &  0.98 &           $..$ &     .. &      .. &      .. &     .. &     .. &    (0.21) &     L &  sp/AO &   X2 &     X* &  .. & large bipolar \\
                 OMN2000\,LS1 & 19234764+1436391 & 4319942771347742976 &  1.24 &           $..$ &   0.91 &    1.27 &    1.15 &     .. &     .. &      1.36 &    SR &      ecl &   .. &      X &   X &               \\
                        GKF2010\,MN107 & 19240333+1339493 & 4319545989397144064 &  1.16 &           $..$ &     .. &      .. &      .. &     .. &     .. &       $..$ &    .. &     .. &   .. &      X &   X &               \\
                        GKF2010\,MN112 & 19443759+2419058 & 2020723052711113344 &  0.99 &           $..$ &   0.20 &    0.18 &    0.20 &     .. &     .. &      0.23 & SDOR: &     .. &   .. &      X &   X &               \\
                          G79.29+0.46 & 20314228+4021591 & 2067716793824240256 &  1.03 &           $..$ &   0.34 &    0.29 &    0.28 &     .. &     .. &       $..$ &    .. &     .. &   .. &      $-$ &   X &               \\
                              MWC\,349A & 20324553+4039366 & 2067727788940487424 &  3.08 &           $..$ &   0.24 &    0.34 &    0.26 &     .. &     .. &      0.14 &  SDOR &     .. &   .. &     X* &   X &               \\

\hline
\end{tabular}
}
\end{sidewaystable*}

\section{Infrared photometry}
\label{infraredtable}
Matches were searched in the 2MASS, MSX, WISE, GLIMPSE, 
and MIPSGAL catalogs  and visually inspected.
The near-infrared Two Micron All Sky Survey (2MASS) survey is described in 
\citet{skrutskie06}, and in \citet{cutri03}.
 74 out of 75 collected stars have a 2MASS counterpart; for  one  star
without 2MASS data, $K$-band magnitudes were given in SIMBAD.

For  51 stars, positional matches within 5\arcsec\ were retained 
in the Midcourse Space Experiment 
catalog \citep[MSX, catalog V/114/msx6\_gp at CDS][]{egan03}, and for  64 
in the WISE survey \citep[catalog II/328/allwise from CDS,][]{wright10}; 
for  58 stars, matches within 4\arcsec\ were found 
in the GLIMPSE catalog (catalog II/293  from CDS) survey \citep{churchwell09} 
and, for  35 stars, in the MIPSGAL 24  \um\ survey 
\citep[catalog J/AJ/149/64  in CDS,][]{gutermuth15}. 
In conclusion,  for all stars, 
there are some available mid-infrared measurements.

Because several targets lie in crowded fields or bright nebulosity, 
all automated infrared matches were inspected visually. 
For DWC2011\,141 for which 2MASS image was confused, 
$HK$ measurements were taken from \citet{dong11}.

Some WISE measurements were  rejected because of confusion --
2MASS\,J17461798$-$2849034, 2MASS\,J17453782$-$2857161, 
2MASS\,J17454015$-$2900276,
2MASS\,J19234764+1436391, 
G0.120-0.048, GKF2010\,MN30, 
2MASS\,J17445538$-$2941284,
DWC2011\,140, LBV 1806$-$20, GKF2010\,MN79, GKF2010\,MN81,
GKF2010\,MN68, GKF2010\,MN76, LBV 1813$-$178,
GKF2010\,MN87, GKF2010\,MN96, DWC2011\,141.

28 stars have 8 \um\ measurements from both the MSX and the GLIMPSE survey;
the mean value of their  $A-[8.0]$
color  is $+0.08$ mag, with an rms of $+0.51$ mag.

For DWC2011\,140, the eclipse binary reported in this work,
we retained the 2MASS photometry 
(\Ks$=7.030\pm0.020$ mag, quality 'AAA'), 
like in the work of \citet{geballe19} and \citet{clark21}.
The 2MASS magnitudes are consistent also
with the measurements of the SIRIUS survey 
\citep[$H=7.6$ mag, $K=7.0$ mag,][]{dong11,nishiyama06}.
\citet{nogueras19} report \Ks $=11.744\pm0.088$ mag. 
However, the star is saturated in the UKIDSS images 
\citep[Kmag1$=11.004\pm0.002$ mag,][]{lucas08} 
and VVV images \citep[Ksmag3$=7.941\pm0.003$ mag,][]{minniti17}).\\
The infrared measurements
of the LBVs and cLBVs are listed in Table \ref{table_phot}.

\begin{sidewaystable*}\renewcommand{\arraystretch}{0.8}
\vspace*{16.5cm} 
\caption{\label{table_phot} Infrared photometry. }
\begin{myresizeenv}
\begin{centering}
\begin{tabular}{@{\extracolsep{-.07in}}l|lrrr|rrrr|rrrr|rrrr|r|}
\hline
\hline
                       Name &   \multicolumn{4}{c}{\rm 2MASS}  &  \multicolumn{4}{c}{\rm WISE}  & \multicolumn{4}{c}{\rm MSX}     &\multicolumn{4}{c}{\rm GLIMPSE}  & MIPSGAL  \\
                            &    2MASS &  J     &  H     &  K     & W1    &  W2    &  W3   &  W4    & A   &  C   &  D   &  E   &3\_6 &  4\_5  &  5\_8  &  8\_0  &  24\_0\\
                            &          &  [mag] &  [mag] &  [mag] & [mag] &  [mag] &  [mag]&  [mag] & [mag] &  [mag] &  [mag] &  [mag] &[mag]&  [mag] &  [mag] &  [mag] &  [mag]\\
\hline
                                $\eta$\,Carinae  & 10450360$-$5941040 &  1.66 &  1.07 &  0.17 &  $..$ &  $..$ &  $..$ &  $..$ & $..$ & $..$ & $..$ & $..$ &   $..$ &   $..$ &   $..$ &   $..$ & $..$ \\
                                P\,Cygni & 20174719+3801585 &  3.62 &  3.41 &  3.05 &  $..$ &  $..$ &   1.72 &   0.81 &  2.06 &  1.59 &  1.49 &  1.06 &   $..$ &   $..$ &   $..$ &   $..$ & $..$ \\
                               HR\,Carinae  & 102253854$-$5937283 &  4.56 &  3.89 &  3.64 &   4.78 &   4.22 &   0.31 &  $-$2.27 &  2.58 &  1.20 &  0.51 & $-$1.38 &   $..$ &   $..$ &   $..$ &   $..$ & $..$ \\
                                AG\,Carinae  & 10561158$-$6027128 &  5.42 &  5.08 &  4.53 &   3.79 &   3.12 &   1.93 &  $-$4.78 &  2.55 &  1.27 &  0.74 & $-$1.96 &   $..$ &   $..$ &   $..$ &   $..$ & $..$ \\
          WRAY\,15-751/IRAS\,11065$-$6026 & 11084006$-$6042516 &  7.96 &  7.24 &  6.75 &   3.90 &   3.43 &   0.72 &  $-$3.32 &  2.92 &  0.77 & $-$0.42 & $-$3.02 &   $..$ &   $..$ &   $..$ &   $..$ & $..$ \\
                          GKM2012 WS1 & 13362862$-$6345387 & 10.33 &  9.48 &  8.89 &   8.03 &   7.60 &   6.80 &   4.58 & $..$ & $..$ & $..$ & $..$ &   $..$ &   $..$ &   $..$ &   $..$ & $..$ \\
                           WRAY\,16-137 & 13501536$-$6148552 &  6.55 &  5.34 &  4.41 &   4.50 &   3.71 &   3.09 &   1.88 &  2.95 &  1.95 &  2.12 &  1.48 &    4.08 &   $..$ &    3.73 &   $..$ &  1.68 \\
                         GKF2010\,MN44 & 16323995$-$4942137 &  8.41 &  7.46 &  6.81 &   5.41 &   4.92 &   4.44 &   0.40 &  6.03 & $..$ &  3.20 &  0.71 &    6.88 &    6.43 &    6.16 &    5.91 & $..$ \\
               Cl* Westerlund 1 W 243 & 16470749$-$4552290 &  6.41 &  5.27 &  4.63 &  $..$ &  $..$ &  $..$ &  $..$ & $..$ & $..$ & $..$ & $..$ &    4.10 &   $..$ &    3.74 &    3.49 &  2.91 \\
                         GKM2012 MN48 & 16493770$-$4535592 &  7.24 &  6.09 &  5.42 &   4.33 &   3.63 &   3.81 &   3.28 &  4.66 & $..$ & $..$ & $..$ &   $..$ &    5.02 &    4.52 &    4.34 &  4.02 \\
                           WRAY\,17-96 & 17413543$-$3006389 &  6.71 &  5.52 &  4.80 &  $..$ &  $..$ &   2.10 &   2.10 &  3.42 &  3.10 &  2.92 & $..$ &   $..$ &   $..$ &   $..$ &   $..$ &  1.29 \\
                            HD\,160529 & 17415902$-$3330136 &  3.55 &  3.06 &  2.79 &  $..$ &  $..$ &   2.22 &   1.79 &  2.29 &  2.19 &  2.11 & $..$ &   $..$ &   $..$ &   $..$ &   $..$ &  1.77 \\
                         G0.120$-$0.048 & 17460562$-$2851319 & 12.53 &  9.24 &  7.46 &  $..$ &  $..$ &  $..$ &  $..$ & $..$ & $..$ & $..$ & $..$ &    5.02 &   $..$ &    4.33 &    4.37 & $..$ \\
                               FMM362 & 17461798$-$2849034 & 12.31 &  8.97 &  7.09 &  $..$ &  $..$ &  $..$ &  $..$ & $..$ & $..$ & $..$ & $..$ &    5.32 &   $..$ &    4.68 &    4.79 & $..$ \\
                              MWC\,930 & 18262523$-$0713177 &  6.65 &  5.81 &  5.26 &   3.91 &   3.27 &   3.01 &   2.01 &  3.74 &  3.32 &  3.18 & $..$ &   $..$ &   $..$ &   $..$ &   $..$ & $..$ \\
                          G24.73+0.69 & 18335528$-$0658386 &  8.36 &  6.84 &  5.92 &   4.15 &   3.50 &   3.56 &   1.81 &  3.68 &  3.21 &  3.14 & $..$ &    4.46 &    4.46 &    3.94 &    4.00 &  3.50 \\
                             AFGL2298 & 19001089+0345471 & 12.16 &  8.92 &  7.01 &   5.62 &   4.42 &  $-$0.56 &  $-$3.50 &  1.85 & $-$0.59 & $-$1.71 & $-$3.75 &    5.20 &   $..$ &   $..$ &   $..$ & $..$ \\
                            SWB2007 1 & 10401937$-$5949099 & 10.55 & 10.18 &  9.95 &   9.77 &   9.56 &   3.90 &  $-$0.15 &  5.71 &  3.25 &  2.57 &  0.34 &   $..$ &   $..$ &   $..$ &   $..$ & $..$ \\
                            Hen\,3-519 & 10535958$-$6026444 &  7.32 &  6.72 &  6.10 &   5.57 &   4.90 &   4.08 &  $-$0.15 &  4.80 & $..$ & $..$ &  1.41 &   $..$ &   $..$ &   $..$ &   $..$ & $..$ \\
                          GKF2010\,MN1 & 11441803$-$6245210 & 10.00 &  8.72 &  7.78 &   6.86 &   6.29 &   5.53 &   3.21 &  5.88 & $..$ & $..$ & $..$ &    6.98 &   $..$ &    6.17 &    5.81 &  4.77 \\
                          WRAY\,15-906 & 11544353$-$6313311 &  7.70 &  6.87 &  6.24 &   5.35 &   4.69 &   3.99 &   2.97 &  4.78 & $..$ & $..$ & $..$ &    7.26 &    4.99 &    4.60 &    4.18 & $..$ \\
                          GKF2010\,MN7 & 13100486$-$6311300 &  9.54 &  7.95 &  6.38 &   4.63 &   3.35 &   1.92 &   0.63 &  2.39 &  1.74 &  1.59 &  1.01 &    4.46 &    4.00 &    3.49 &   $..$ &  0.97 \\
                          GKF2010\,MN8 & 13104384$-$6317457 & 12.27 & 11.00 & 10.19 &   9.45 &   8.95 &   8.52 &   4.75 & $..$ & $..$ & $..$ & $..$ &    9.52 &    9.06 &    8.82 &    8.45 &  7.63 \\
                         GKF2010\,MN13 & 13423308$-$6248113 & $..$ & 13.27 & 10.07 &   6.36 &   4.58 &   1.68 &  $-$0.69 &  2.87 &  1.60 &  1.32 & $-$0.02 &    6.84 &    6.00 &    4.33 &   $..$ & $..$ \\
                         GKF2010\,MN30 & 15484207$-$5507422 &  7.93 &  6.72 &  5.98 &  $..$ &  $..$ &   4.78 &   4.20 &  4.51 & $..$ & $..$ & $..$ &   $..$ &    5.14 &    4.94 &    4.78 &  3.92 \\
                         GKF2010\,MN39 & 16102654$-$5121253 & $..$ & $..$ & 11.26 &   8.18 &   6.31 &   6.17 &   3.06 & $..$ & $..$ & $..$ & $..$ &    7.02 &    6.00 &    5.35 &    5.29 &  4.91 \\
             IRAS\,16115$-$5044 & 16151795-5052197 &  7.76 &  6.13 &  5.14 &   5.42 &   4.78 &   2.36 &  $-$4.80 & $..$ & $..$ & $..$ & $..$ &   $..$ &    3.81 &    3.69 &    3.32 & $..$ \\
                         GKF2010\,MN41 & 16263427$-$5021018 &  9.93 &  8.55 &  7.74 &   5.65 &   5.05 &   4.53 &  $-$0.14 &  6.08 & $..$ &  2.83 & $-$0.28 &    6.52 &    6.16 &    5.55 &    5.36 & $..$ \\
       WMD2010\,14/IRAS\,16254$-$4739 & 16290377-4746264 &  8.67 &  7.14 &  6.19 &   4.70 &   4.20 &   3.99 &   0.04 &  4.08 &  3.30 &  3.15 &  1.08 &    6.75 &    5.63 &    5.08 &    4.74 & $..$ \\
                         GKF2010\,MN42 & 16313781$-$4814553 & $..$ & 12.60 & 10.28 &   8.53 &   7.53 &   6.58 &   1.62 & $..$ & $..$ & $..$ & $..$ &    8.17 &    7.46 &    7.11 &    6.85 & $..$ \\
                         GKF2010\,MN45 & 16364278$-$4656207 & 11.71 &  7.89 &  5.82 &   5.40 &   4.12 &   4.32 &   0.44 &  3.99 &  3.79 &  3.60 &  1.38 &    4.92 &    4.52 &    4.05 &    4.05 &  4.00 \\
                GKF2010\,MN46 & 16431636-4600424 &  6.26 &  5.08 &  4.21 &  $..$ &  $..$ &  $..$ &  $..$ &  2.41 &  1.55 &  1.06 &  0.23 &   $..$ &   $..$ &   $..$ &   $..$ &  0.22 \\
                         GKF2010\,MN47 & 16461734$-$4508478 & 15.38 & 11.87 &  9.85 &   7.90 &   6.99 &   6.35 &   1.88 & $..$ & $..$ & $..$ & $..$ &    8.09 &    7.56 &    7.25 &    7.15 & $..$ \\
                            HD\,326823 & 17065390$-$4236397 &  6.71 &  6.10 &  5.45 &   4.53 &   3.51 &   2.83 &   1.99 &  3.13 &  2.49 &  2.44 & $..$ &    4.36 &    3.91 &    3.70 &   $..$ & $..$ \\
GKF2010\,MN51 & 17082913-3925076 & 10.49 &  8.22 &  7.11 &   6.36 &   6.09 &   5.16 &   3.97 &  5.25 & $..$ & $..$ & $..$ &    6.46 &    6.16 &    5.71 &    5.54 &  3.95 \\
                         GKF2010\,MN53 & 17092477$-$4008455 & $..$ & 10.84 &  8.53 &   6.72 &   5.71 &   5.38 &   3.92 &  5.41 & $..$ & $..$ & $..$ &    6.75 &    6.39 &    5.45 &    5.19 &  4.14 \\
                         GKF2010\,MN55 & 17110094$-$3945174 &  9.73 &  8.04 &  6.96 &   6.01 &   5.27 &   4.70 &   1.73 &  4.88 & $..$ & $..$ & $..$ &    6.67 &    6.08 &    5.24 &    4.88 &  3.67 \\
                         GKF2010\,MN56 & 17111668$-$3931312 & 10.47 &  8.77 &  7.83 &   6.60 &   5.31 &   0.48 &  $-$2.24 &  2.06 &  1.23 &  0.86 & $-$1.76 &   $..$ &    6.77 &   $..$ &   $..$ & $..$ \\
                          GKM2012\,WS2 & 17203175$-$3309490 &  6.62 &  5.41 &  4.70 &   4.07 &   3.16 &   2.64 &   1.63 & $..$ & $..$ & $..$ & $..$ &   $..$ &   $..$ &   $..$ &   $..$ & $..$ \\
                         GKF2010\,MN58 & 17374754$-$3137333 & 11.77 &  9.25 &  7.79 &   6.29 &   5.86 &   5.79 &   3.89 &  6.00 & $..$ & $..$ & $..$ &    6.84 &    6.19 &    5.96 &    5.83 &  5.03 \\
                         GKF2010\,MN61 & 17421401$-$2955360 & 13.38 & 10.54 &  8.99 &  $..$ &  $..$ &  $..$ &  $..$ &  5.32 & $..$ & $..$ & $..$ &    7.68 &    7.16 &    6.84 &    6.86 & $..$ \\
               2MASS\,J17444319$-$2937526 & 17444319$-$2937526 & $..$ & 12.78 & 10.14 &   8.12 &   6.71 &   4.88 &   3.33 &  5.03 &  3.72 & $..$ & $..$ &    7.62 &    6.78 &    6.07 &    5.67 &  3.21 \\
                         GKF2010 MN64 & 17435981$-$3028384 &  5.74 &  4.78 &  4.07 &   3.29 &   2.65 &   1.56 &  $-$0.55 &  2.88 &  1.39 &  0.85 & $-$0.31 &    4.45 &    3.61 &    3.52 &   $..$ & $..$ \\
              2MASS\,J17445538$-$2941284 & 17445538$-$2941284 & 16.72 & 12.63 & 10.14 &   7.79 &   6.63 &  $..$ &  $..$ & $..$ & $..$ & $..$ & $..$ &    7.74 &    7.04 &    6.49 &    6.19 & $..$ \\
                 DWC2011\,140 & 17445945$-$2905258 &  8.71 &  7.59 &  7.03 &   6.19 &   6.03 &  $..$ &  $..$ & $..$ & $..$ & $..$ & $..$ &    6.78 &    6.21 &    5.90 &    5.79 &  4.71 \\
              2MASS\,J17450241$-$2854392 & 17450241$-$2854392 & 15.81 & 12.56 & 10.00 &   6.77 &   5.20 &   3.34 &   1.92 &  3.57 &  2.59 &  2.65 &  1.94 &    6.86 &    6.00 &    4.79 &    4.25 &  2.16 \\

\hline
\end{tabular} 
\end{centering}
\end{myresizeenv}
\end{sidewaystable*}

\addtocounter{table}{-1}
\begin{sidewaystable*}\renewcommand{\arraystretch}{0.8}
\vspace*{16.5cm} 
\caption{Continuation of \ref{table_phot}. }
\begin{myresizeenv}
\begin{centering}
\begin{tabular}{@{\extracolsep{-.07in}}l|lrrr|rrrr|rrrr|rrrr|r|}
\hline
\hline
                       Name &   \multicolumn{4}{c}{\rm 2MASS}  &  \multicolumn{4}{c}{\rm WISE}  & \multicolumn{4}{c}{\rm MSX}     &\multicolumn{4}{c}{\rm GLIMPSE}  & MIPSGAL  \\
                            &    2MASS &  J     &  H     &  K     & W1    &  W2    &  W3   &  W4    & A   &  C   &  D   &  E   &3\_6 &  4\_5  &  5\_8  &  8\_0  &  24\_0\\
                            &          &  [mag] &  [mag] &  [mag] & [mag] &  [mag] &  [mag]&  [mag] & [mag] &  [mag] &  [mag] &  [mag] &[mag]&  [mag] &  [mag] &  [mag] &  [mag]\\
\hline
                          DWC2011\,141 &             $..$ & $..$ & 15.30 & 11.70 &  $..$ &   6.11 &   4.51 &  $..$ &  4.82 &  3.87 &  3.37 &  2.77 &    7.63 &    6.48 &    5.46 &    4.87 &  2.69 \\
                          DWC2011\,112 & 17453782$-$2857161 & 15.55 & 12.30 & 10.29 &  $..$ &  $..$ &  $..$ &  $..$ & $..$ & $..$ & $..$ & $..$ &    7.95 &    7.15 &    6.30 &    5.56 & $..$ \\
                           Pistol Star      & 17461524$-$2850035 & 11.83 &  8.92 &  7.29 &  $..$ &  $..$ &  $..$ &  $..$ & $..$ & $..$ & $..$ & $..$ &    6.67 &    6.00 &    5.06 &    4.97 & $..$ \\
              2MASS\,J17470940$-$2849235 & 17470940$-$2849235 & 12.66 & 10.80 &  9.46 &   7.77 &   7.15 &   4.96 &   3.26 &  5.74 &  4.86 &  4.33 &  2.77 &    7.73 &    7.18 &    6.55 &    5.49 &  2.90 \\
                            HD\,316285 & 17481403$-$2800531 &  4.82 &  4.23 &  3.71 &  $..$ &  $..$ &   1.44 &   0.57 &  1.89 &  1.33 &  1.08 &  0.70 &   $..$ &   $..$ &   $..$ &   $..$ &  0.76 \\
              2MASS\,J17482472$-$2824313 & 17482472$-$2824313 & 15.86 & 12.12 &  9.54 &   7.39 &   6.06 &   3.60 &   0.44 &  4.29 &  3.06 &  2.47 &  0.73 &    7.02 &    6.24 &    5.03 &    4.30 &  0.52 \\
                         GKF2010\,MN68 & 18022233$-$2238002 & 13.60 & 11.07 &  9.60 &   8.46 &   7.73 &  $..$ &   3.65 & $..$ & $..$ & $..$ & $..$ &    8.37 &    7.76 &    7.50 &    7.29 & $..$ \\
                         GKF2010\,MN76 & 18070516$-$2015163 & 14.87 & 12.67 & 11.16 &  10.05 &   9.48 &  $..$ &   5.21 & $..$ & $..$ & $..$ & $..$ &    9.91 &    9.43 &    9.15 &    8.99 & $..$ \\
                          LBV 1806-20 & 18084031$-$2024411 & 13.66 & 10.48 &  8.74 &   6.38 &   5.62 &  $..$ &  $..$ & $..$ & $..$ & $..$ & $..$ &    7.26 &    6.73 &    6.43 &    6.30 & $..$ \\
                           MFD2008\,15 & 18132098$-$1749469 &  8.93 &  7.78 &  6.96 &   6.13 &   5.76 &   5.34 &  $..$ &  5.46 & $..$ & $..$ & $..$ &    6.76 &    5.91 &    5.71 &    5.42 &  4.37 \\
                         GKF2010\,MN79 & 18283340$-$1146441 & 14.36 & 11.42 &  9.71 &   8.34 &   7.42 &  $..$ &  $..$ & $..$ & $..$ & $..$ & $..$ &    8.29 &    7.58 &    7.28 &    7.07 & $..$ \\
                         GKF2010\,MN80 & 18333954$-$0807084 & 13.00 &  9.85 &  8.15 &   6.29 &   6.21 &   6.09 &   3.04 &  6.04 & $..$ & $..$ & $..$ &    7.10 &    6.39 &    6.11 &    6.05 &  5.71 \\
                         GKF2010\,MN81 & 18334346$-$0823353 & $..$ & 12.03 &  9.81 &   8.51 &   7.66 &  $..$ &  $..$ & $..$ & $..$ & $..$ & $..$ &    8.14 &    7.64 &    7.24 &    7.14 & $..$ \\
            SSTGLMC G023.4521-00.0948 & 18342170$-$0828209 &  9.78 &  8.42 &  7.63 &   6.75 &   6.41 &   5.75 &   4.62 &  6.01 & $..$ & $..$ & $..$ &    6.89 &    6.51 &    6.17 &    5.93 &  4.67 \\
                            G25.5+0.2 & 18370520$-$0629379 & $..$ & 13.64 & 11.62 &  $..$ &  $..$ &  $..$ &  $..$ &  4.49 &  1.21 &  0.27 & $-$1.95 &   $..$ &   $..$ &   $..$ &   $..$ & $..$ \\
                         GKF2010\,MN83 & 18392300$-$0553199 & 13.19 & 10.09 &  8.39 &  $..$ &  $..$ &  $..$ &  $..$ & $..$ & $..$ & $..$ & $..$ &    7.09 &    6.62 &    6.26 &    6.24 & $..$ \\
                          G26.47+0.02 & 18393224$-$0544204 &  8.00 &  6.53 &  5.61 &  $..$ &  $..$ &  $..$ &  $..$ &  3.53 &  2.18 &  2.50 & $..$ &    4.80 &    4.21 &    3.81 &   $..$ &  1.49 \\
                         GKF2010\,MN84 & 18415965$-$0515409 &  7.96 &  6.53 &  5.68 &   6.00 &   5.67 &   5.32 &   1.34 &  5.42 & $..$ & $..$ & $..$ &   $..$ &   $..$ &    4.61 &    4.35 &  2.93 \\
                         GKF2010\,MN87 & 18422247$-$0504300 & $..$ & 11.10 &  8.68 &   6.42 &   5.58 &   5.64 &  $..$ & $..$ & $..$ & $..$ & $..$ &    6.87 &   $..$ &    5.75 &    5.67 & $..$ \\
                       IRAS\,18433$-$0228 & 18455526$-$0225089 & $..$ & 14.20 & 11.68 &   9.28 &   8.10 &   4.30 &  $-$0.24 &  6.81 & $..$ &  2.70 &  0.37 &    8.56 &    7.87 &    7.34 &    7.63 &  0.15 \\
                         GKF2010\,MN90 & 18455593$-$0308297 & 15.45 & 11.51 &  9.37 &   7.52 &   6.57 &   5.83 &   1.24 &  6.16 & $..$ & $..$ & $..$ &    7.39 &    6.72 &    6.32 &    6.10 & $..$ \\
                         GKF2010\,MN96 & 18510295$-$0058242 & 11.88 & 10.14 &  9.13 &   8.04 &   7.59 &   7.38 &  $..$ & $..$ & $..$ & $..$ & $..$ &    8.15 &    7.72 &    7.48 &    7.17 &  5.67 \\
                        GKF2010\,MN101 & 19062457+0822015 &  9.66 &  7.92 &  6.85 &   5.75 &   5.04 &   4.21 &   0.60 &  4.47 &  3.40 &  3.31 &  1.45 &    5.80 &    5.43 &    5.10 &    4.80 &  2.56 \\
                              MWC\,314 & 19213397+1452570 &  6.09 &  5.54 &  5.02 &   4.71 &   3.95 &   3.32 &   2.26 &  3.55 &  2.92 &  2.54 & $..$ &    4.89 &    4.61 &    4.07 &    3.77 &  2.12 \\
                 OMN2000\,LS1 & 19234764+1436391 &  8.10 &  7.02 &  6.23 &   6.70 &   6.17 &   4.72 &  $..$ & $..$ & $..$ & $..$ & $..$ &    6.66 &    5.67 &    5.14 &    4.77 & $..$ \\
                        GKF2010\,MN107 & 19240333+1339493 & 10.16 &  9.21 &  8.60 &   8.05 &   7.70 &   7.17 &   5.00 & $..$ & $..$ & $..$ & $..$ &    7.95 &    7.76 &    7.55 &    7.30 &  6.51 \\
                        GKF2010\,MN112 & 19443759+2419058 &  8.86 &  8.02 &  7.42 &   6.75 &   6.28 &   5.56 &   4.11 &  5.93 & $..$ & $..$ & $..$ &    6.73 &    6.42 &    6.18 &    5.84 &  4.80 \\
                          G79.29+0.46 & 20314228+4021591 &  6.91 &  5.29 &  4.33 &   3.71 &   2.97 &   2.77 &   2.47 &  2.78 &  2.47 &  2.13 & $..$ &   $..$ &   $..$ &   $..$ &   $..$ & $..$ \\
                              MWC\,349A & 20324553+4039366 &  6.22 &  4.75 &  3.15 &  $..$ &  $..$ &  $-$1.19 &  $-$2.58 & $-$1.12 & $-$1.93 & $-$2.16 & $-$2.63 &   $..$ &   $..$ &   $..$ &   $..$ & $..$ \\

\hline
\end{tabular} 
\end{centering}
\end{myresizeenv}
\end{sidewaystable*}
\end{appendix}

\end{document}